# Optoelectronic switch based on intrinsic dual Schottky diodes in ambipolar MoSe$_2$ field-effect transistors

*Nihar R. Pradhan,*[1,*] *Zhengguang Lu,*[1] *Daniel Rhodes,*[1] *Dmitry Smirnov,*[1] *Efstratios Manousakis,*[1,2,3] *and Luis Balicas*[1,*]

Dr. N. R. Pradhan, Z. Lu, D. Rhodes, Dr. D. Smirnov, Prof. E. Manousakis, Dr. L. Balicas

National High Magnetic Field Lab, Florida State University, 1800 E. Paul Dirac Drive, Tallahassee, Florida 32310, United States.
E-mail: balicas@magnet.fsu.edu,
E-Mail: pradhan@magnet.fsu.edu

Prof. E. Manousakis
Florida State University, Department of Physics, Tallahassee, FL 32306, United States,

Prof. E. Manousakis
Department of Physics, University of Athens, Panepistimioupolis, Zografos, GR-157 84 Athens, Greece.

Here, we report the observation of a hitherto unreported optoelectronic effect, namely a light-induced diode-like response in multi-layered MoSe$_2$ field-effect transistors whose sense of current rectification is controllable through a gate voltage. We argue, through numerical simulations, that this behavior results from the difference in the size of the Schottky barriers between drain and source metal contacts. Each barrier can be modeled as a Schottky diode but with opposite senses of current rectification between them, with the diode response resulting from the light induced promotion of photo-generated carriers across the smaller barrier. The back gate voltage controls the sense of current rectification by modulating the relative amplitude between them.  This effect, which gives rise to a novel type of optoelectronic switch, also yields a photovoltaic response. Hence, it could provide an alternative to PN-junctions when harvesting photovoltaic currents from transition metal dichalcogenides. We argue that the photovoltaic efficiency associated to this effect could be increased by just increasing the relative asymmetry between both Schottky barriers. We also suggest that this new electro-optical effect has potential for technological applications.

**1. Introduction**

Transition metal dichalcogenides (TMDs) are claimed to absorb between 5 and 10% of the incident sunlight when exfoliated into thicknesses inferior to 1 nm displaying one order of magnitude higher sunlight absorption than most of the commonly used solar absorbers.[1] In monolayers this leads to the observation of unique optical[2,3,4] and optoelectronic properties.[5,6] Pronounced photoconducting[7] and photovoltaic responses are also observed in heterostructures incorporating graphene and multi-layered TMDs.[6,8]

The high light absorption in TMDs was attributed to the existence of pronounced van Hove singularities in the electronic density of states leading to a pronounced joint density of states in the visible light region and hence ensuring relatively strong light-matter interactions.[6,9,10] Strong light matter interactions led to reports of incredibly high photoresponsivities in single-layer $MoS_2$, i.e. approaching $\sim 10^3$ A/W in the limit of very low illumination power densities.[7] For large area, chemical vapor deposited heterostructures of graphene onto $MoS_2$ mono-layers photoresponsivities as high as $10^7$ A/W were reported [11] under illumination power densities $p$ approaching just $\sim 10^{-3}$ W/m$^2$. Very high photoresponsivities and concomitantly high external quantum efficiencies are also claimed for graphene and transition metal dichalcogenides based heterostructures [6,8] even when transferred onto flexible substrates.[6] These heterostructures were reported to display considerably high photovoltaic currents.[8]

In order to further evaluate the suitability of transition metal dichalcogenides for electronics and optoelectronics recent research efforts focused on the fabrication of *p-n* junctions.[12-18] Conventional *p–n* junctions are composed of two regions, a hole or *p*-doped region adjacent to an *n*-doped one with a depleted zone in between, i.e., an area with no charge carriers which is characterized by a pronounced built-in electrostatic potential associated with the uncompensated dopant atoms. Once the charge carriers are accelerated by the built-in electrostatic potential, determined by the spatial extent of the depletion region, the transport across the junction occurs through diffusion and/or drift processes.

To produce *p-n* junctions based on transition metal dichalcogenides, insofar three strategies were reported: i) to vertically stack[12-14] of *p*- (e.g. WSe$_2$) and *n*-doped (e.g. MoS$_2$) single atomic layers in order to create a vertical *p-n* junction and ii) to electrostatically accumulate electrons and holes in contiguous but spatially separated regions[15-17] in an ambipolar transition-metal dichalcogenide single-atomic layer such as WSe$_2$ and iii) the use of asymmetric contacts, or the use of distinct metals characterized by different work functions for either contact, in combination with the chemical doping of a portion of the conduction channel.[14] All of these approaches led to well-defined diode like *I-V* characteristics with diode ideality factors *n* ranging between $n = 1$ and 2.[8-12]

Upon illumination all three approaches led to sizeable so-called short circuit currents or currents in absence of a bias voltage resulting from the photovoltaic effect. Nevertheless, photovoltaic power conversion efficiencies η were reported only in Ref.[13], for vertical junctions yielding η = 0.2 %, and in Refs.[16,17] for lateral junctions yielding η = 0.5 and 0.14 %, respectively. These values were extracted assuming that the active photovoltaic area is just the depletion area of the junction, and not the entire illuminated area of the conduction channel. Obviously, when compared to the current photovoltaic technologies these efficiencies are still small for technological applications. In addition to the photovoltaic response, vertical[14] as well lateral[15-18] *p-n* junctions were also reported to display electroluminescence, hence holding a certain potential for light emitting display applications.

To harvest a photovoltaic response from thin layers of transition metal dichalcogenides, other approaches were also explored. For example, large area monolayer $MoS_2$ synthesized by chemical vapor deposition and directly transferred onto *p*-Si, when contacted with an Al wire array, is claimed to yield η = 5.23 %.[19] Vertical heterostructures composed of Au as the top electrical contact and a multilayered $MoS_2$ crystal on an indium tin oxide (ITO) substrate, which is characterized by a sizeable Schottky barrier between $MoS_2$ and the Au contact, was shown to exhibit η up to 1.8 %.[20] On the other hand, vertical heterostructures composed of an ITO layer as the top contact, Au as the bottom contact, and two $MoS_2$ crystals which are respectively electron- and hole-doped through a plasma induced process yield η = 2.8 %.[21] Finally, the use of two metals with distinct work functions, such as Pd and Au for the drain and source current contacts respectively, are expected to pin the Fermi level at distinct positions at either contact. This creates a gradient of the chemical potential, which upon illumination, was recently shown to also yield a photovoltaic response with a maximum power conversion efficiency η = 2.5 %.[22]

In this manuscript we report the observation of a back-gate tunable/controlled light induced diode-like response in few layered $MoSe_2$ field-effect transistors contact with Ti:Au. We observe photovoltaic response in the absence of an obvious PN junction in the fabricated device. This response can be significantly altered by applying a back-gate potential and can be well-described by the Schockley-diode equations yielding a diode ideality factor smaller than 2. More importantly, we find that the application of a back-gate voltage can change the direction of the photocurrent. We argue that this behavior emerges from the presence of two effective Schottky diodes induced at the metallic source and drain contacts. The back-gate voltage modifies the effective height of the Schottky barriers on both the source and the drain contacts and changes the direction of the induced net photocurrent according to its sign. We describe this behavior through numerical simulations considering two such Schottky diodes connected in series characterized by opposite sense of current rectification. In addition, we consider i) an asymmetry in the size of the Schottky barriers at each current contact, ii) the simultaneous role of bias and gate voltages, as well as iii) the role of light as the generator of electron-hole pairs. At each current contact surface states pin the Fermi level at different distinct positions relative to the bands of MoSe$_2$, generating a gradient of the chemical potential across the conducting channel which leads to the experimental observation of a finite photovoltaic current in the absence of a bias voltage. We also argue that the application of a gate-voltage contributes to the vertical spatial separation between electrons and holes increasing their recombination times, and hence the concomitant photovoltaic current. Our observations indicate that the higher the asymmetry in the size of the respective Schottky barriers, the more pronounced the diode-like response and hence the higher the extracted photovoltaic power conversion efficiency. We argue that the photovoltaic efficiencies extracted from such simple devices could be increased simply by increasing the asymmetry between the sizes of the Schottky barriers at each contact or by increasing the number of MoSe$_2$ layers. Furthermore, we argue that the back-gate can be used as a control nob to alter the magnitude and direction of the photo-current which might make this a potentially useful device for optical communications.

## 2. Results and Discussion

**Figure 1a** displays a micrograph of a seven-layer MoSe$_2$ field-effect transistor (sample#1) prepared for electrical transport measurements, including Hall-effect and Hall mobilities,[23] and which was used for the photo-current study presented here. Throughout this manuscript we will focus only on two-terminal measurements, where both the voltage and the electrical current are read and biased through the drain ($I^+$) and the source ($I^-$) contacts. All measurements shown in here were collected at room temperature. The current voltage characteristics for both samples whose data is shown here, as well as a scheme of the experimental set-up used for the electrical transport measurements under illumination can be found in the Supplementary Information file, see Figs. S1 and S2 respectively. Figure 1b shows a height profile collected along the edge of the crystal indicating a thickness of ~ 5 nm or approximately 7 atomic layers. Figure 1c displays the drain to source current $I_{ds}$ as a function of the gate voltage $V_{bg}$ under a bias voltage $V_{ds} = 0.1$ V, and for several temperatures. As seen, when contacted with Ti:Au, $MoSe_2$ displays ambipolar behavior, yielding a sizeable $I_{ds}$ for both positive (accumulation of electrons in the channel) and negative (accumulation of holes) gate voltages. For this particular sample one extracts far more current for electrons accumulated in the channel when compared to the holes particularly at lower temperatures. Figure 1d displays the respective field-effect mobilities $\mu_{FE} = 1/c_g \, d\sigma/dV_{bg}$ as a function of the temperature $T$, where $c_g = \varepsilon_r \varepsilon_0/d = 12.789 \times 10^{-9}$ F/cm$^2$ is the gate capacitance for a $d = 270$ nm thick $SiO_2$ layer, and $\sigma = l \, I_{ds}/wV_{ds}$ is the conductivity for a channel of width $w = 6.7$ μm and length $l = 9.4$ μm. Here, $d\sigma/dV_{bg}$ was extracted from a simple linear fit of $I_{ds}(V_{bg})$ taken at high values of $I_{ds}$. As seen, although the electron field-effect mobility (or $\mu^e_{FE}$ which is depicted by orange markers) is observed to increase and saturate, the hole-mobilities (or $\mu^h_{FE}$ and depicted by blue markers) decrease very fast as $T$ is lowered. The red line is a linear fit yielding $\mu^e_{FE} \propto T^{-1.22}$ which most likely reflects the suppression of phonon scattering. For this particular sample, $\mu^e_{FE}$ saturates at a value of ~ 600 cm$^2$/Vs at low temperatures. Notice also that the observed low temperature values are higher than those extracted for multi-layered $MoS_2$ (see Ref. [24]) or for multi-layered $WSe_2$ contacted with graphene and gated by using an ionic liquid (see Ref. [25]). We note that in Ref.[23] we reported $MoSe_2$ samples displaying considerably higher Hall and field-effect mobilities at room temperature, indicating that it is possible to achieve considerably higher electron mobilities at low temperatures then the ones reported here.

**Figure 2a** depicts the main observation within this manuscript: a plot of $I_{ds}$ as a function of the bias voltage $V_{ds}$ under zero gate-voltage, respectively under dark conditions (magenta-line) and under illumination (blue-line). An illumination power $P = 30$ µW was applied to the sample by a $\lambda$=532 nm laser through an optical fibre placed in very close proximity to its surface producing a spot size of 3.5 µm in diameter. Although the *I-V* characteristics collected under dark conditions is already asymmetric or non-linear (as indicated by the inset which plots $I_{ds}(V_{bg}=0$ V, $P=0$ µW) as a function of $V_{ds}$) as expectable for non-ohmic contacts, it becomes far more asymmetric, and in fact diode-like *under illumination*. As shown in Figure 2b, rectification like behaviour becomes even more pronounced when a gate voltage $V_{bg} = +7.5$ V is applied. Remarkably, the sense of current rectification can be inverted by just inverting the gate voltage as shown in Fig. 2c which displays $I_{ds}$ as a function of $V_{bg}$ under $P = 30$ µW for $V_{bg} = + 7.5$ and -7.5 V, respectively. Red and brown lines are fits to Fits of the observed diode response to the Shockley equation in the presence of a series resistor:[26]

$$I_{ds} = \frac{nV_T}{R_s} W_0\left(\frac{I_0 R_s}{nV_T} \exp\left(\frac{V + I_0 R_s}{nV_T}\right)\right) - I_0 \qquad (1)$$

where $W$ is the Lambert function and $V_T$ the thermal voltage, yields an ideality factor $n \cong 1.1$ with the series resistance $R_s$ ranging from 0.85 to 2.1 MΩ and $I_0$ from 5 x $10^{-11}$ and 2 x $10^{-9}$ A. These *n* values are smaller than those of Refs.[15,16] ($1.9 \leq n \leq 2.6$) for single layered WSe$_2$ lateral diodes. In any case, the Shockley-Read-Hall recombination theory,[27,28] which assumes recombination via isolated point defect levels, predicts $n \leq 2$. Here, it is important to emphasize that the area of the channel $A_c = 6.7$ µm x 9.4 µm $\cong 63$ µm² is much larger than the area of the laser spot or $A_l = \pi(3.5/2)^2 \cong 9.6$ µm², implying that the illumination of the channel is responsible for the observed effect, since we were able to easily prevent illumination of the contacts. We have previously found that illumination of the contacts does affect the optoelectronic response of transition metal dichalcogenides when contacted with metallic electrodes, see Ref. [29] and related supporting information.

Therefore, the application of illumination power leads to a striking diode-like response in an otherwise field-effect transistor, whose sense of current rectification can be controlled by the gate voltage. Since this response is observed in the absence of a conventional p-n junction it can only be ascribed to the Schottky barriers at the level of the contacts in the spirit of the results and samples described in Refs.[20,22] This will be discussed in detail below.

In **Figure 3** illustrates that a photovoltaic response can be extracted from a dual Schottky barrier inherent to $MoSe_2$ field-effect transistor in absence of a gate voltage. Figure 3a displays the $I_{ds}$ as a function of $V_{ds}$ characteristics for a *second* transistor, or sample #2, under $V_{bg}$ = 0 V and several values of the laser illumination power $P$. As seen, the response is non-linear which, as we will discuss below, can be attributed to the Schottky barriers between $MoSe_2$ and the electrical contacts. Notice how for this sample, in contrast to sample # 1, the *I-V* characteristics are far more symmetric with respect to the sign of the bias voltage. Through numerical simulations, we will argue below that its higher symmetry with respect to sample #1 is attributable to a smaller difference in the height of both Schottky barriers. In the Supplementary Information file we provide a detailed electrical characterization under illumination power, showing that this sample also displays a light-induced, gate voltage controllable rectification response, see **Figure S3**. Figure 3b plots $I_{ds}$ as a function of $V_{bg}$, same data as in Figure 3a but in an amplified scale, where one can observe the short-circuit current, or the photo-generated current in the absence of a bias voltage, or the photovoltaic effect. In the same figure one can extract the photo-generated voltage in the absence of any electrical current or the open circuit voltage $V_{oc}$. Figure 3c displays the concomitant electrical power $P_{el} = I_{ds}$ x $V_{ds}$ within the quadrant spanning from $I_{ds}$ = 0 A to $V_{ds}$ = 0 V, which is attributable solely to the photovoltaic effect. This data was collected from sample #2. As seen, in the absence of any gate voltage one can still extract some very small electrical power, in the order of 100 pW, implying very small power conversion efficiencies. The important point is that the collection of charge carriers through the contacts, and in the absence of a conventional PN-junction or of a bias voltage, is possible in this system being consistent with a gradient of the chemical potential. This would be due, for example, to Fermi surface pinning by impurities around the contact area, or asymmetric Schottky barriers, which would accelerate carriers towards the contacts. It could also be attributed to an illumination induced thermal gradient between the channel and the contacts.

**Figure 4a** displays $I_{ds}$ as a function of $V_{ds}$, now for sample #1, but under $V_{bg}$ = + 7.5 V and for several values of the illumination power illustrating again that a photovoltaic response can be extracted from such simple devices**.** Figure 4b shows again the same data but in an amplified scale in order to expose both $I_{sc}$ and $V_{oc}$. Figure 4c plots the corresponding photo generated electrical power $P_{el}$. Notice how under $V_{bg}$ = +7.5 V and for the same $P$s, one ends up extracting lower $P_{el}$ values than what is extracted for sample #2 under $V_{bg}$ = 0 V. In contrast, and as seen in Figure 4d**,** one can obtain far more pronounced photocurrents under $V_{bg}$ = - 7.5 V (larger by more than one order of magnitude when compared to $V_{bg}$ = +7.5 V) thus indicating a very pronounced asymmetry with respect to the sign of the gate voltage. Even more pronounced are the concomitant values for $I_{sc}$ and $V_{oc}$ as shown in Figure 4e. As seen in Figure 4f**,** and by comparing it to panel 4c, one ends up extracting far more pronounced $P_{el}$ values indicating that a higher asymmetry leads to a more pronounced photovoltaic response. In the Supplementary Information file, we have included plots of the photocurrent and of the open circuit voltage as a function of the applied optical power in **Figure S5**. We have also included photoresponsivities and concomitant external quantum efficiencies for this sample, see **Figure S6**, which yield maximum values, for this experimental configuration approaching, or exceeding, 100 mA/W and 30 %, respectively.

Here, we propose that our field-effect transistors are controlled by the Schottky barriers[30] between the Ti/Au metallic contacts and the $MoSe_2$ layers as illustrated by Figure 5 which as we discuss below can be modeled as two Schottky diodes with opposite sense of current rectification. **Figure 5a** depicts a sketch of a multi-layered MoSe$_2$ field-effect transistor contacted with Ti:Au under laser illumination and an applied back-gate voltage. The gate voltage spatially separates photo-generated electron and holes towards the bottom and the top layer which leads to an increase in the electron-hole recombination times. The sense of rectification is simply defined by the geometry of the sample with the Au contacts located to the right and to the left of the MoSe$_2$ channel at the source and drain contacts, respectively. The presence of states at the metal/MoSe$_2$ interface, such as dangling bonds, inevitable impurities,[31] and perhaps even residues of the lithographic process,[30] pin the Fermi level somewhere in the middle of the semiconducting gap as suggested by Bardeen.[32] Hence, both Schottky barriers (and concomitant Schottky diodes) would not be identical and, in general, they would be expected to have some small difference Δ between the respective barrier heights as depicted in Figure 5b. In general we find that it is almost impossible to fabricate absolutely identical electrical contacts in transition metal dichalcogenides. We insist that even for bulk samples, e.g. intermetallics, organic conductors, transition metal oxides etc., in our extensive experience the electrical contacts are never absolutely identical even when using conventional methods such as silver epoxies or silver paints. Slight difference in contact geometry can also affect their resistance. However, the application of a gate voltage modulates or reduces the size of both barriers as illustrated by green lines in Figure 5b. But as we discuss below, our simulations and concomitant fits indicate that $V_{bg}$ can affect/reduce the height of each barrier by different fractions. As illustrated by Figures 5c and 5d, the simultaneous application of light which separates/generates electrons and holes and of a bias voltage which tilts the valence and the conduction bands, allows one to harvest of either type of charge carrier. The kind of carrier to be harvested depends on the sign of the gate voltage which displaces the Fermi level towards either the conduction or the valence band. The sense of current rectification on the other hand depends on the relative size between both Schottky barriers, as defined by $V_{bg}$. At the moment we do not have a clear understanding on the role of the gate voltage in modulating the relative asymmetry between both Schottky barriers. Our study indicates that this effect cannot be observed in strongly asymmetric barriers, e.g. when distinct metals are used for the drain and source contacts.

In a circuit model these two diodes are in series and they have opposing orientation for their forward biases as schematically shown in the Figure 6a. Furthermore, under illumination they produce photo-generated currents flowing in opposite directions (represented by red arrows) as is discussed below. Representation of a single Schottky diode by means of a Schottky diode current and a photo-generated current flowing in the opposite direction as illustrated in Figure 6a is standard.[33,34] In the case of our devices, as we will demonstrate below, the photovoltaic response is due to a dual Schottky diode configuration. The circuit that captures the essential aspects of our device is illustrated by Figure 6a.

We have used two Schottky diodes in series to represent the two opposite rectification Schottky diodes formed at the metal-semiconductor (source) and semiconductor-metal (drain) contacts. The resistor $R_S$ represents the intrinsic resistance of the device. We use the standard approach of using the Shockley diode equation to approximate each Schottky diode.[33,34] The Kirchhoff-Shockley equations of this circuit are the following:

$$I_{ds} = I_L^{(s)} - I_D^{(s)} = I_D^{(d)} - I_L^{(d)}, \qquad (2)$$

$$V_{ds} = V_{AB} + V_{BC} + I_{ds} R_S, \qquad (3)$$

$$I_D^{(s)} = I_0^{(s)} \left( \exp\left(\frac{V_{BA}}{n_s \kappa_B T}\right) - 1 \right), \qquad (4)$$

$$I_D^{(d)} = I_0^{(d)} \left( \exp\left(\frac{V_{BC}}{n_d \kappa_B T}\right) - 1 \right), \qquad (5)$$

where, $I_D^{(s,d)}$ is the source and drain diode currents and $I_L^{(s,d)}$ the photo-generated currents; $n_s$ and $n_d$ are the ideality factors of each of the two Schottky diodes. The potential differences, i.e., $V_{AB}$ and $V_{BC}$ are indicated in Figure 5a. For given values of the parameters $I_0^{(s,d)}, I_L^{(s,d)}, R_S$ and $n_s$, $n_d$, the above sets of equations can be solved self-consistent (using the Newton-Raphson technique) to obtain the $I_{ds}$ versus $V_{ds}$ characteristics. Typical solutions of the above equations are illustrated by the fittings in Figure 6b and 6c (dashed red lines). The results fit reasonably well the experimental data by using ideality factors $n_s$ and $n_d$ close to unity. The only difference in the extracted parameters obtained by fitting the *I-V* characteristics under various illumination powers *P*, including *P* = 0 W, and for any given sample, are in the values of the photocurrent parameters $I_L^{(s)}$ and $I_L^{(d)}$ (see the Supplementary Information file for tables containing the extracted parameters). The other parameters remain independent of *P*, which indicates that the physics of the device is correctly captured by the above simple model. Furthermore, the values of $I_L^{(s)}$ and $I_L^{(d)}$ scale roughly linearly in *P*. We find that the values of $I_L^{(s)}$ and $I_L^{(d)}$ are strongly dependent on the asymmetry of the Schottky barriers which is markedly sample dependent. For example, sample #1 is far more asymmetric than sample #2.

We also find that the application of a back-gate voltage can alter the relative size of both Schottky barriers, i.e. they are not necessarily displaced in energy by the exact same amount as depicted in Figure 4b. For photovoltaic applications, we observe that the size of the short circuit current $I_{sc}$ could be modulated simply by altering the relative size between both Schottky barriers by tuning the back-gate voltage. For example, for sample # 1 the relative size of the Schottky barriers favored current rectification along one specific sense. However, by inverting the gate voltage we were able to alter their relative size and allow current rectification in the opposite sense. By increasing the carrier recombination times due to the vertical separation of carriers by the gate voltage one can considerably increase the short circuit current.

If one was able to construct a device with very asymmetric Schottky barriers, e.g. a very large barrier in one contact and a nearly ohmic second contact, one would be able to harvest a very sizeable short-circuit current due to the pronounced gradient of the chemical potential due to the bending of the bands. Furthermore, as shown here one might be able to: i) further increase the short circuit current with the application of a back-gate voltage and, ii) control the sense of circulation of the short circuit current with the sign and amplitude of the back-gate voltage. Nevertheless, we find that a very pronounced asymmetry leads to just one sense of current rectification. Such a pronounced asymmetry on the other hand, is likely to produce a sizeable photovoltaic response in the absence of doping or of a more complex architecture such as a PN junction. This is in fact suggested by the overall photovoltaic evaluation presented in the Supplementary **Figure S6**.

## 3. Conclusion

In summary, a light-induced diode-like response is observed in multi-layered $MoSe_2$ field-effect transistors electrically contacted with Ti:Au. Our numerical simulations indicate that this behavior results from an asymmetry in the size of the Schottky barriers between drain and source contacts with each barrier being described as a Schottky diode but with opposite senses of current rectification. This diode-like response would result from the light induced promotion of photo-generated carriers across the smaller between both barriers. Remarkably, the sense of current rectification can be effectively controlled by the back-gate voltage through its ability to modulate the relative amplitude between both barriers. The precise role played by the gate voltage on the height of the respective Schottky barriers will require further detailed studies. This diode response yields a photovoltaic response in the absence of a PN-junction such as those created by either electrostatic or chemical doping opening up the possibility of fabricating simpler and cost effective solar cells based on transition metal dichalcogenides. By learning how to precisely control the size of the Schottky barriers between transition metal dichalcogenides and the metallic contacts, it should be possible to fabricate field-effect transistors displaying a more pronounced diode like response upon illumination. Their asymmetry in height already leads to photovoltaic power conversion, and the extracted conversion efficiencies might increase by just increasing their relative asymmetry, e.g. by combining a large work function metal such as Pd for one the contacts, with a small work function one such as Sc for the other. Although Schottky barriers are widely believed to limit the performance of electronics and optoelectronics based on transition metal dichalcogenides, our results indicate they can also offer additional and potentially useful functionalities.

Our observations indicate that transition metal dichalcogenides based field-effect transistors can be used as new type of photo-switches. This would correspond to a hitherto unreported electro-optic phenomenon. In its usual connotation, the electro-optic effect refers to a change in the optical properties of any given material upon the application of a slowly varying electric field. Typical examples include changes in the absorption of light such as the Franz-Keldish[35] and the quantum-confined Stark effect[36] or changes in the refractive index such as the Pockels[37] and the Kerr[38] effects. Relevant technologies have emerged from their discovery, hence suggesting that the effect described here might have an interesting potential for technological applications. Here, the technological challenge is to precisely control the height of the Schottky barriers upon device fabrication and to understand in detail, how their relative height evolves upon application of a gate voltage. Following Ref.[39] which demonstrated the feasibility of logical inverters and logical NOR operations in dual gated field-effect transistors, one could envision similar architectures based on these optically activated diodes to produce logical operations. In this case, one could perform logical operations by controlling their respective senses of current rectification with gate voltages, while possessing the ability of activating the logical elements with an optical signal.

## 4. Experimental Section

$MoSe_2$ single-crystals were synthesized through a chemical vapor transport technique using iodine as the transport agent. Multi-layered flakes of $MoSe_2$ were exfoliated from these single-crystals by using the micromechanical exfoliation technique, and transferred onto *p*-doped Si wafers covered with a 270 nm thick layer of $SiO_2$. For making the electrical contacts 90 nm of Au was deposited onto a 4 nm layer of Ti *via* e-beam evaporation. Contacts were patterned using standard e-beam lithography techniques. After gold deposition, these devices were annealed at 250° C for ~ 2 h in forming gas. This was followed by a subsequent high vacuum annealing for 24 h at 120°C. Atomic force microscopy (AFM) imaging was performed using the Asylum Research MFP-3D AFM. Electrical characterization was performed by using Keithley 2612 A sourcemeter. Measurements as a function of the temperature were performed in a Physical Property Measurement System. The sample was kept under a low pressure of $^4$He as exchange gas. For photo-current measurements a Coherent Sapphire 532-150 CW CDRH and Thorlabs DLS146-101S were used, with a continuous wavelength of 532 nm. Light was transmitted to the sample through a 3 μm single-mode optical-fiber with a mode field diameter of 3.5 μm. The size of the laser spot was also measured against a fine grid. Hence, here we use 3.5 μm for the laser spot diameter assuming a constant power density distribution in order to approximate the Gaussian distribution corresponding to the mode field diameter of 3.5 μm. Particular care was taken to avoid illumination of the electrical contacts, although it is likely that some scattered or reflected light illuminated the area contiguous to the contacts. For the dark current measurements, the device under test was covered with a black box (with controlled access for the laser beam) to prevent any exposure to light.

**Supporting information**

Following supporting information is available from the Wiley Online Library. Current as a function of the bias voltage, description of the experimental set-up, light induced diode response from a second field-effect transistor, photocurrent, open-circuit voltage, photoresponsivity, external quantum efficiencies and overall evaluation of the photovoltaic response. We also include the table with the parameters extracted from the simulations. Correspondence and requests for materials should be addressed to N.R.P., E.M. and L.B.

**CORRESPONDING AUTHORS**
Corresponding authors
*E-mail: pradhan@magnet.fsu.edu
*E-mail: balicas@magnet.fsu.edu

**ACKNOWLEDGMENTS**
This work is supported by the U.S. Army Research Office MURI grant W911NF-11-1-0362. ZL, and DS acknowledge the U.S. Department of Energy, Office of Basic Energy (DE-FG02-07ER46451) for the support for the photo-response measurements. The NHMFL is supported by NSF through NSF-DMR-0084173 and the State of Florida.

Received: ((will be filled in by the editorial staff))
Revised: ((will be filled in by the editorial staff))
Published online: ((will be filled in by the editorial staff))

**Figure 1.**

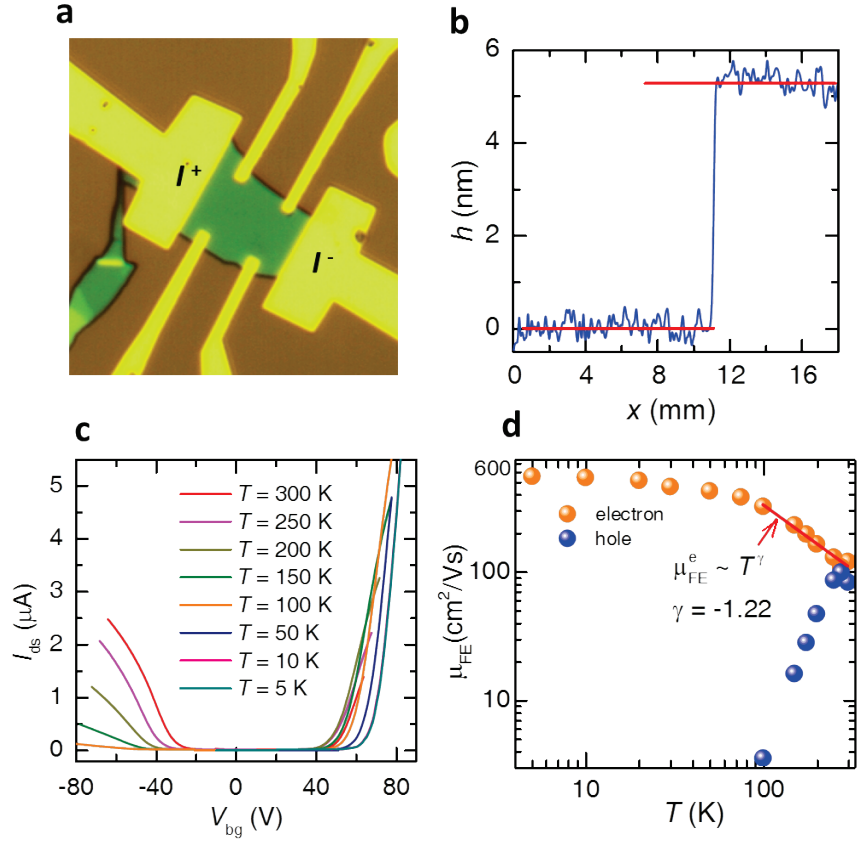

**Figure 1.** a) Micrograph of one of our MoSe$_2$ field-effect transistors on SiO$_2$. Drain ($I^+$) and source ($I^-$) current contacts are indicated. b) Height profile as extracted from atomic force microscopy across the edge of the crystal indicating a thickness of ~ 5 nm, which divided by the interlayer spacing c = 6.4655 Å, yields ~ 7 atomic layers. c) Drain to source current $I_{ds}$ as a function of the back-gate voltage $V_{bg}$ under a bias voltage $V_{ds}$ = 0.1 V and for several temperatures ranging from 300 to 5 K. d) Field effect mobility $\mu_{FE}$= 1/$c_g$ d$\sigma$/ d$V_{bg}$ as a function of the temperature $T$, where $c_g = \varepsilon_r\varepsilon_0/d$ = 12.789 x 10$^{-9}$ F/cm$^2$ is the gate capacitance for a $d$ = 270 nm thick SiO$_2$ layer, and $\sigma = l\, I_{ds}/wV_{ds}$ is the conductivity for a channel of width $w$ (= 6.7 μm) and length $l$ (= 9.4 μm). The electron mobility $\mu^e_{FE}$ is observed to increase by a factor of ~ 5 as the $T$ is lowered while the hole-mobility decreases very rapidly. This indicates either a considerable increase in the threshold gate voltage for hole-conduction upon cooling, or the suppression of thermionic hole-emission across the Schottky barriers at the contacts. Red line is a linear fit yielding $\mu^e_{FE} \sim T^{1.22}$.

**Figure 2.**

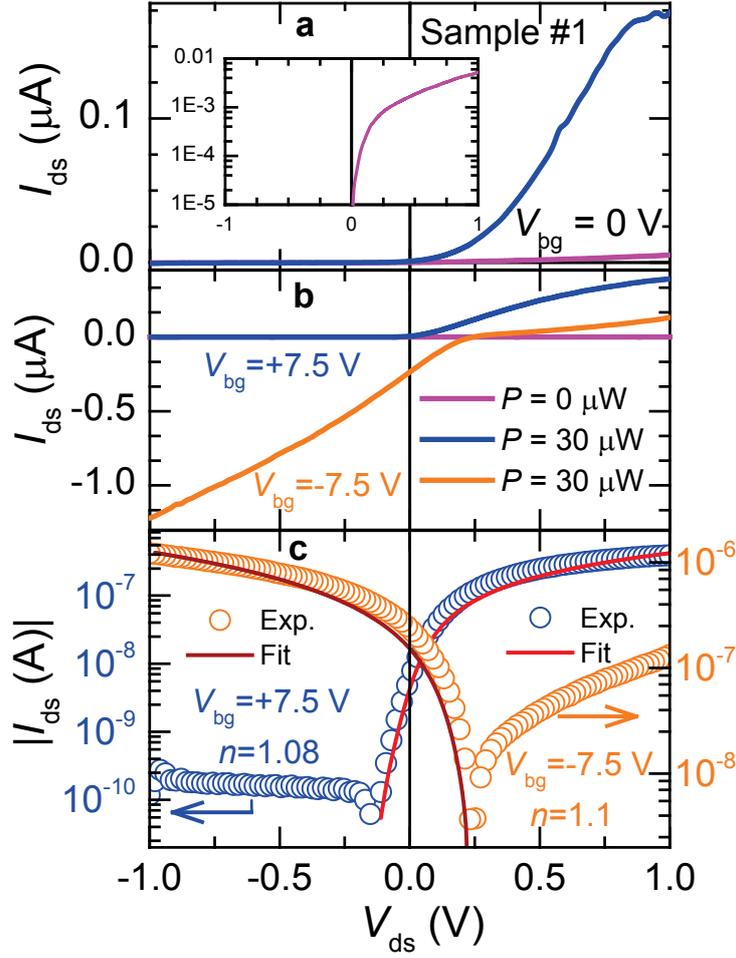

**Figure 2.** a) $I_{ds}$ as a function of the bias voltage $V_{ds}$ under zero gate voltage ($V_{bg}$ = 0 V) for the MoSe$_2$ field-effect transistor shown in Figure 1, under respectively 0 (magenta line) and 30 µW (blue line) of illumination power $P$. Inset: $I_{ds}$ in a logarithmic scale as a function of $V_{ds}$ under 0 µW. As seen the *I-V* characteristics is non-linear, indicating a prominent role for the Schottky barriers at the electrical contacts. b) Same as in a) but under two values of the gate voltage, $V_{bg}$ = + 7.5 V (blue markers) and – 7.5 V (orange markers), respectively. Notice how the diode-like response induced by illumination becomes far more pronounced upon the application of a gate voltage, which can also control the sense of current rectification. c) Absolute value of $I_{ds}$ in a logarithmic scale and as a function of $V_{ds}$ for both values of $V_{bg}$. Notice how the sense of current rectification can be controlled by the polarity of the gate voltage. Red and brown lines are fits to the Shockley diode equation including a series resistance $R_s$ which yields diode ideality factors $n \cong 1.1$
21

**Figure 3.**

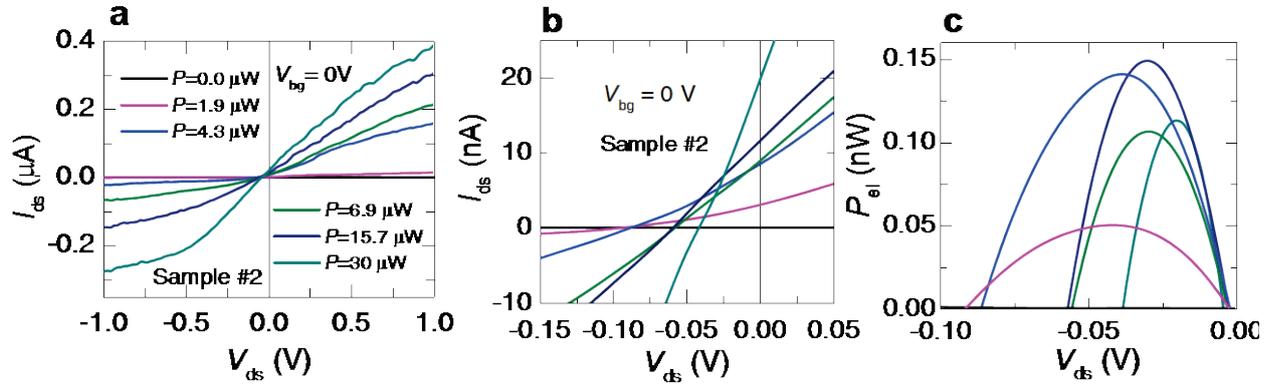

**Figure 3.** a) $I_{ds}$ as a function of $V_{ds}$ for several values of the applied illumination power $P$ for a second sample, or sample #2, of similar thickness. The non-linear component of the *I-V* characteristics can be simulated by two Schottky diodes with rectification currents flowing in opposite directions as discussed in the main text. b) Same as in a) but in an amplified scale. Notice the finite current in absence of any excitation voltage, or the short circuit current $I_{sc}$ resulting from the photovoltaic effect. c) Photo-generated electrical power $P_{el}$ as a function of $V_{ds}$.

**Figure 4.**

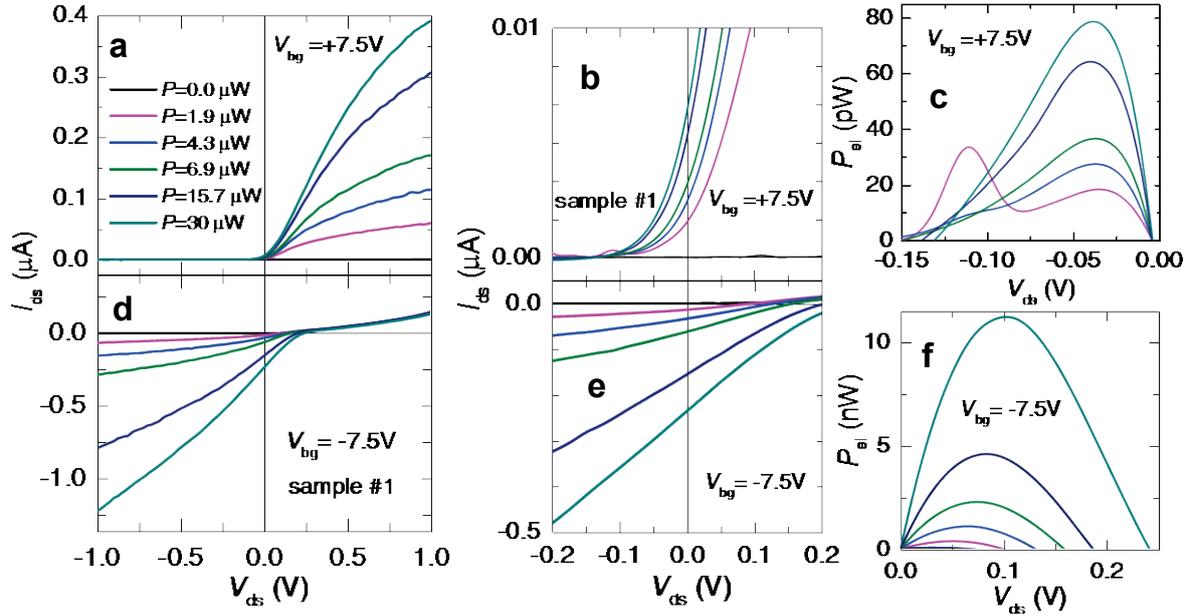

**Figure 4.** a) $I_{ds}$ as a function of $V_{ds}$ for sample #1, and for several values of the illumination power $P$ under $V_{bg}$ = + 7.5 V. b) Same as in a) but in an amplified scale to expose the short circuit current $I_{sc}$ = $I_{ds}$ ($V_{ds}$ = 0 V). (c) Photo-generated electrical power $P_{el} = I_{ds} \times V_{ds}$ as a function of $V_{ds}$ from the data in b). d) Same as in a) but under $V_{bg}$ = - 7.5 V ($P$ is indicated by line colors). e) Same as in d) but in an amplified scale to expose $I_{sc}$. f) $P_{el}$ as a function of $V_{ds}$ as extracted from the data in e).

**Figure 5.**

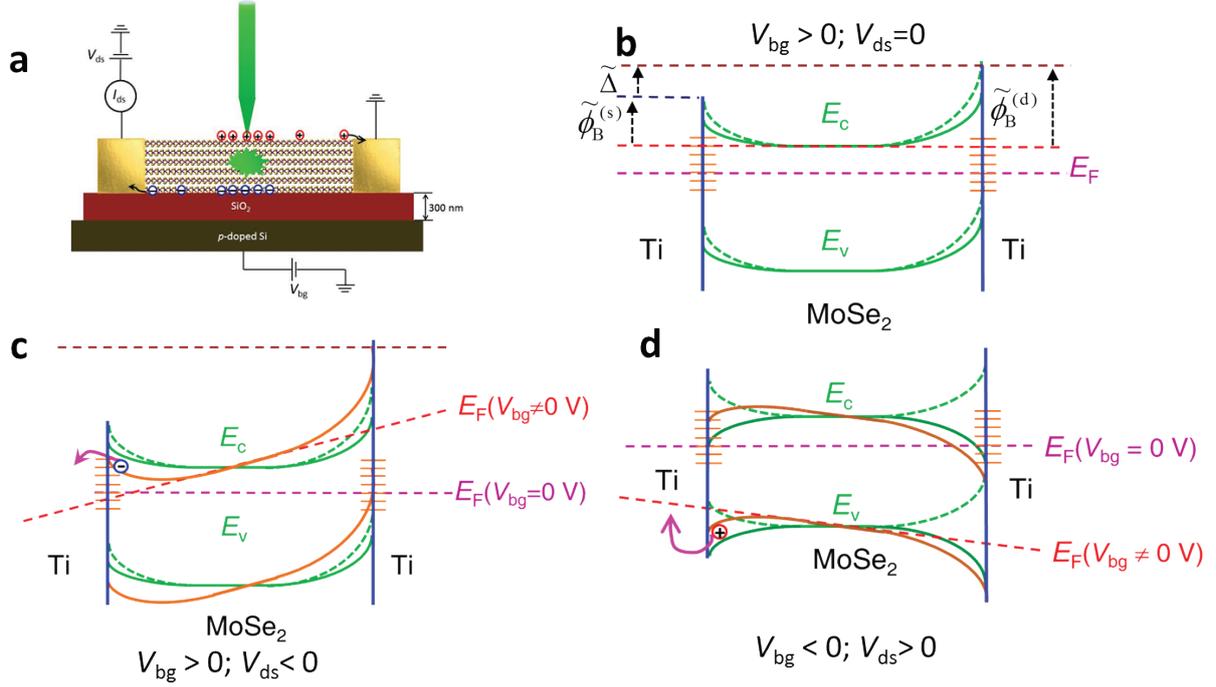

**Figure 5.** a) Sketch of a MoSe$_2$ field-effect transistor under illumination and under applied gate and excitation voltages. The photogenerated electron-hole pairs are spatially separated by the gate voltage which increases considerably the characteristic exciton recombination times. b) Effective band scheme for our samples assuming non-identical contacts exhibiting a difference $\Delta$ in the size of their respective Schottky barriers $\phi_B$ between MoSe$_2$ and the Ti contacts. Green-dashed lines depict $E_v$ and $E_c$ which stand for valence and conduction bands, respectively while $E_F$ (depicted by the purple line) stands for the Fermi level in *absence* of gate-voltage. The presence of states at the metal/MoSe$_2$ interface, such as dangling bonds, etc., pin the Fermi level somewhere in the middle of the semiconducting gap as suggested by Bardeen.[31] These mid gap states are depicted by horizontal orange lines. Hence, the situation depicted here favors the flow of current towards the source contact. The application of a gate voltage modulates the amplitude of both Schottky barriers and can decrease their size by turning for example, the barrier at the source $\phi_B^{(s)}$ into $\widetilde{\phi}_B^{(s)}$. Continuous lines depict the valence and conduction bands relative to the Fermi level $E_F$ under a back-gate voltage $V_{bg}$. Their curvature in the neighborhood of the contacts as well as their position with respect to $E_F$ is controlled by $V_{bg}$. Horizontal red line depicts the displaced Fermi level when a gate voltage is applied. Green-dashed lines indicate the original $E_v$ and $E_c$ before the application of $V_{bg}$. Here, we depict a similar gate-voltage induced reduction for both Schottky barriers although our experimental results indicate that the size of each barrier is reduced by a distinct relative amount which is dependent upon the polarity and amplitude of the gate-voltage. c) Same as in a) but where the orange line depicts the profile of the conduction and valence bands once a bias voltage is applied. In this cartoon the scheme of both bands and their relative position with respect to the Fermi level $E_F$ would allow the collection of photo-generated electrons. d) The application of a negative gate voltage displaces the Fermi level towards the valence band, while the application of a bias voltage of opposite polarity (with respect to c)) would allow the collection of photo-generated holes. Our simulations indicate that the gate voltage reverses the asymmetry between Schottky barriers. The combination of gate and bias voltages, as well as the relative asymmetry between both Schottky barriers (controlled by the gate voltage), leads to the observed rectification-effect and determines the type of carrier to be harvested.

**Figure 6.**

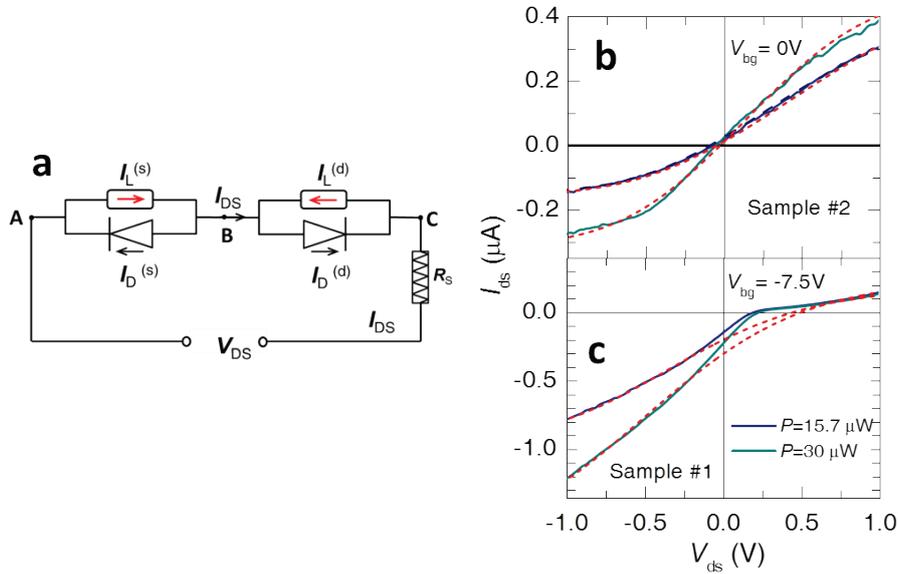

**Figure 6.** a) Equivalent circuit which we believe captures the observed behavior. Each contact is characterized by a Schottky diode whose rectification current flows in the opposite sense with respect to one another. Identical diodes would lead to no net photocurrent under illumination. But the application of a gate voltage modulates the relative amplitude between both barriers, allowing the photo generated current (depicted here as current sources and/or red arrows) to flow in either direction depending on the sign of the gate voltage. The resistor $R_s$ represents a shunt resistance such as the resistance of the channel. b) Fits of experimental data from sample # 2 under $V_{bg}$ = 0 V and two values of the illumination power $P_{opt}$, 15.7 µW and 30 µW, to the dual Schottky circuit model in a), and which are depicted by red dashed lines. c) Same as in b) but for sample #1 under $V_{bg}$ = -7.5 V and for the same $P$ values.

**TOC Figure.**

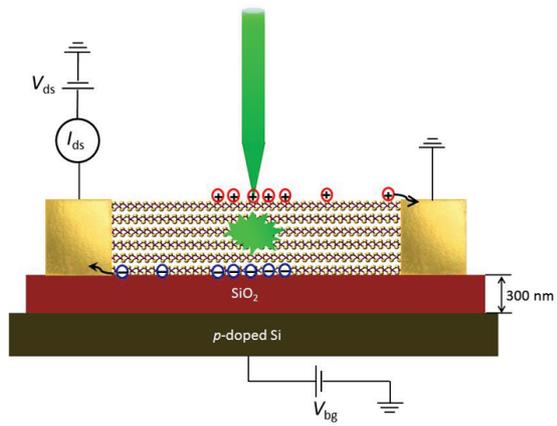 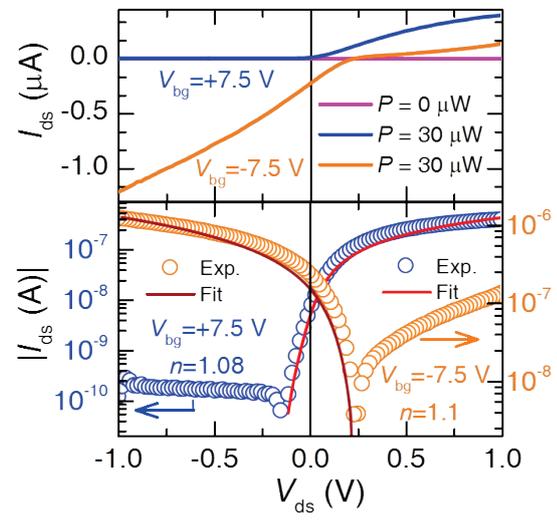

Supporting Information

# Optoelectronic switch based on intrinsic dual Schottky diodes in ambipolar MoSe$_2$ field-effect transistors

*Nihar R. Pradhan,*[*] *Zhengguang Lu, Daniel Rhodes, Dmitry Smirnov, Efstratios Manousakis,*[*]

*and Luis Balicas*[*]

Dr. N. R. Pradhan, Z. Lu, D. Rhodes, Dr. D. Smirnov, Prof. E. Manousakis, Dr. L. Balicas
National High Magnetic Field Lab, Florida State University, 1800 E. Paul Dirac Drive, Tallahassee, Florida 32310, United States.
E-mail: balicas@magnet.fsu.edu,
E-Mail: pradhan@magnet.fsu.edu

Prof. E. Manousakis
Florida State University, Department of Physics, Tallahassee, FL 32306, United States,

Prof. E. Manousakis
Department of Physics, University of Athens, Panepistimioupolis, Zografos, GR-157 84 Athens, Greece.

1. **Current as a function of bias voltage characteristics for both MoSe$_2$ field effect transistors.**

**Figure S1** displays the *I-V* characteristics of samples #1 and #2 under several back voltages and under dark conditions, revealing a pronounced non-linear *I-V* response from sample #1, particularly at low excitation voltages. This non-ohmic behavior can be attributed to pronounced Schottky barriers between the MoSe$_2$ channel and the Au:Ti contacts. In contrast, sample #2 presents a nearly linear, or ohmic like response at low bias voltages. This indicates that thermionic emission can promote charge carriers across the barriers at room temperature.

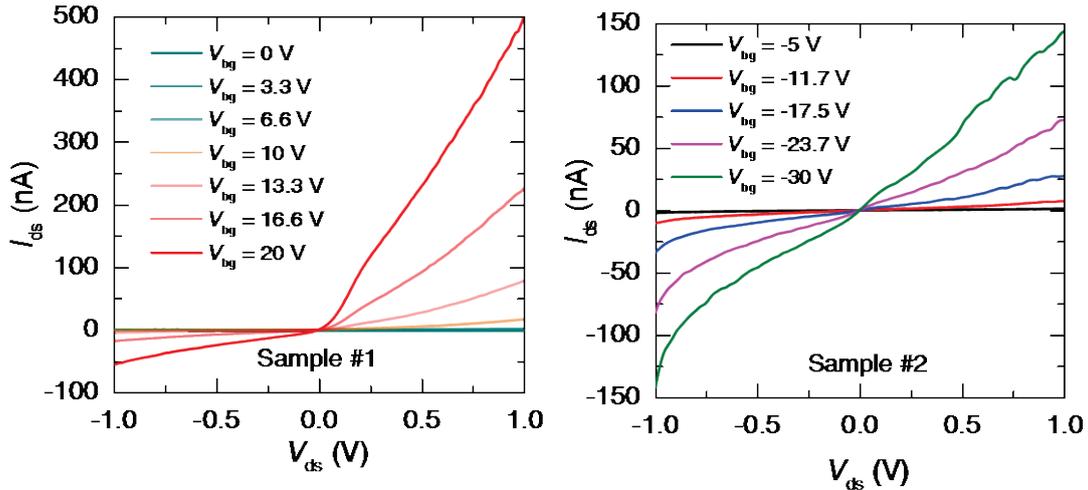

**Figure S1.** Left panel: Drain to source current $I_{ds}$ as a function of the bias voltage $V_{ds}$ for sample #1 and for several values of back gate voltage $V_{bg}$. Right panel: same as in the left panel but for sample #2.

## 2. Experimental set-up for measuring the electrical response under illumination power.

**Figure S2** below presents a sketch of the experimental set up used for the electrical characterization of our samples under dark or illumination conditions. The diode laser beam is sent to the sample through an optical fiber, subsequently collimated, and focused through an achromatic objective. If necessary its intensity can be modulated by using neutral density filters. The sample is measured through a dual channel sourcemeter.

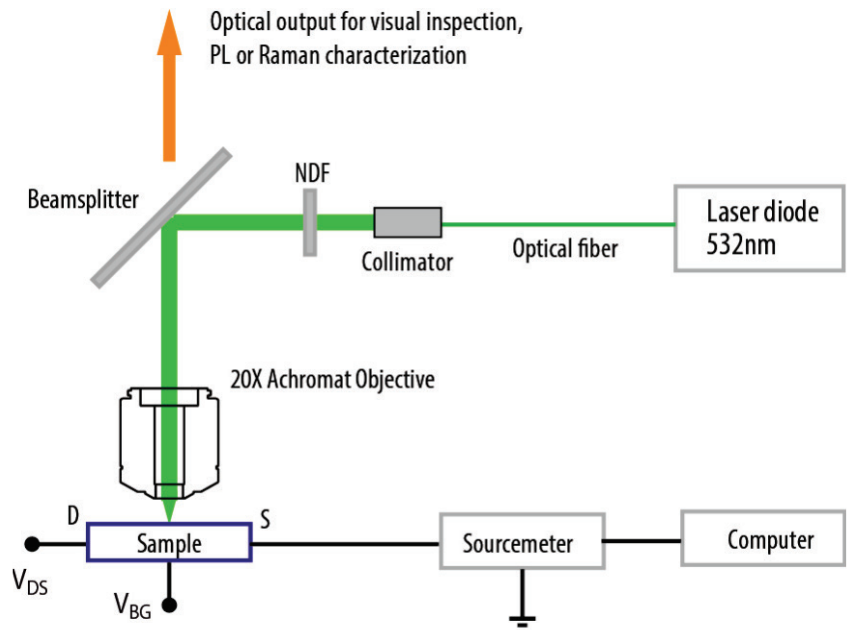

**Figure S2.** Schematics of the experimental set-up used to measure the electrical response of our MoSe$_2$ transistors under optical stimulation.

## 3. Light induced rectification response in a second multi-layered and a tri-layered MoSe$_2$ field-effect transistor

**Figure S3** below displays the overall photo response from a second MoSe$_2$ field-effect transistor (sample #2 in the main text) under illumination by a $\lambda$ = 532 nm laser with a spot size of 3.5 μm.

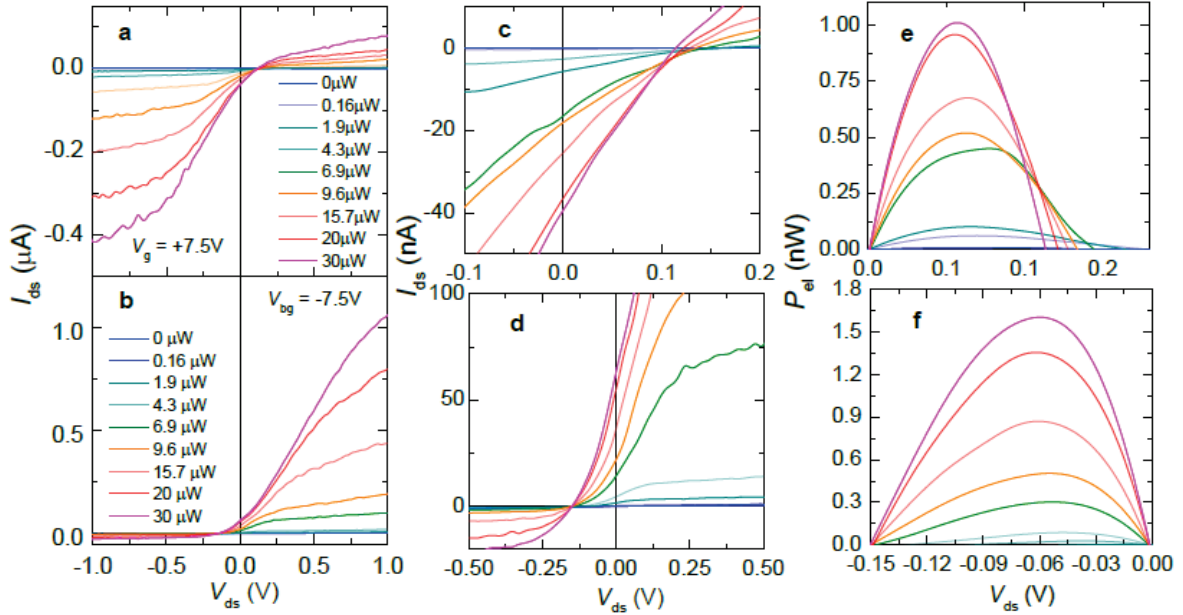

**Figure S3.** a) Drain to source current $I_{ds}$ as a function of the bias voltage $V_{ds}$ for several values of the illumination power and under a gate voltage $V_{bg}$ = - 7.5 V. b) Same as in a) but under $V_{bg}$ = - 7.5 V. The line color scheme which associated to the laser power is the same for all figures. c) Same as in a) but in an amplified scale. d) Same as in b) but in an amplified scale. Notice, by comparing c) and d), how under both gate voltages one obtains relatively similar values for the short circuit currents $I_{sc}$ = $I_{ds}(V_{ds}$ = 0 V). e) Photo-generated electrical power $P_{el}$ = $I_{ds}$ x $V_{ds}$ as extracted from the fourth quadrant, spanning from $V_{ds}$ = 0 V to $I_{ds}$ = 0 A, in c). f) same as in e) but from the data in d). Notice how in sharp contrast to sample #1, sample # 2 yields comparable $P_{el}$ values for both gate voltages although this higher degree of symmetry leads to much lower photogenerated power values when compared to the ones extracted from sample #1 under $V_{bg}$ = - 7.5 V.

Notice how for this sample by inverting the sign of the gate voltage one can also switch the sense of current rectification. Although, with respect to sample #1, in this sample the sense of current rectification displays an opposite dependence on the sign of the gate voltage. For sample #1 one extracts a sizeable $I_{ds}$ only when *both* $V_{ds}$ and $V_{bg}$ are either > 0 or < 0 in contrast to what is seen here. Again, this indicates that the gate voltage is modulating the amplitude of each Schottky barrier in an independent and non-trivial manner, perhaps by pinning the Fermi level at distinct impurity levels at each contact. Notice that for sample #2 under zero bias and for both gate voltages one extracts similar values for the short circuit current suggesting that for this sample the gate voltage can invert the assymmetry between both Schottky barriers. It also points towards a lower degree of asymmetry in the height of both barriers. As seen, a higher degree of symmetry leads to lower electrical power conversion levels, when compared for example, to sample #1 under $V_{bg} = -7.5$ V (see Figure 4 f in the main text). For the sake of comparison with the data displayed in the main text (from a multi-layered sample) in **Figure S4** below we display data from a tri-layered sample, or sample #3, indicating that the rectification effect is also observed in thinner samples characterized by a larger, and nearly direct semiconducting gap.

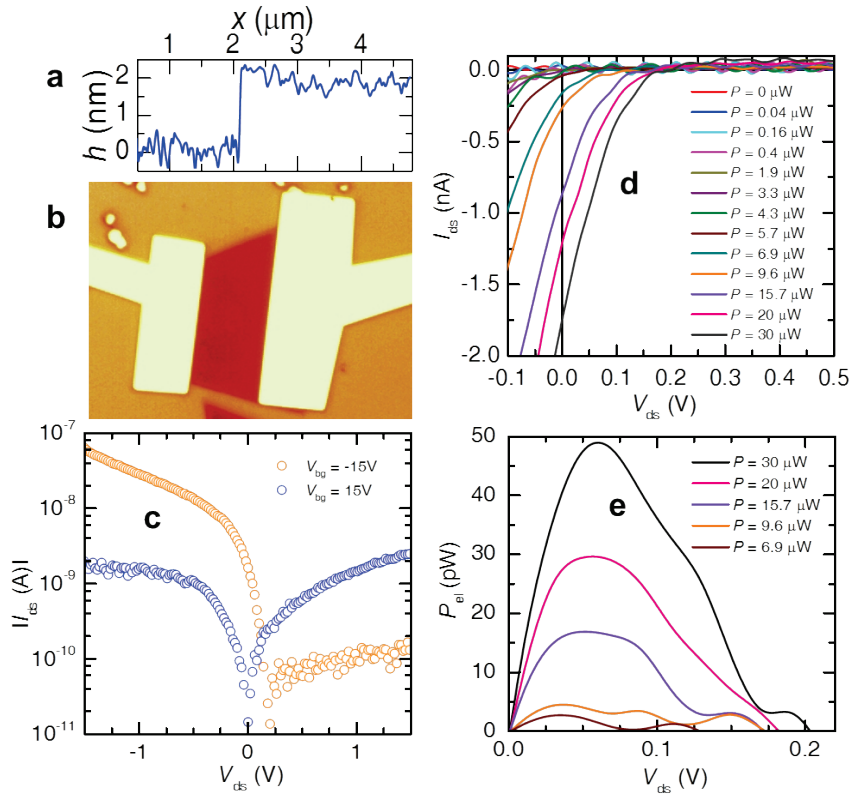

**Figure S4.** a) Atomic force microscopy height profile of a third sample, i.e. sample #3, indicating a thickness of approximately 2 nm or 3 atomic layers. b) Micrograph of sample#3. c) Drain to source current $I_{ds}$ as a function of the bias voltage $V_{ds}$ for two values of the gate voltage, i.e. respectively 15 and -15 V under an illumination power of $P = 30$ μW. d) $I_{ds}$ as a function of $V_{ds}$ for several values of $P$. Notice the finite photogenerated current in absence of bias voltage. e) Photogenerated electrical power $P_{el} = I_{ds} \times V_{ds}$ as a function $V_{ds}$ for several values of the illumination power.

As seen in Figure S4e the maximum $P_{el}$ output power and more generally the response of sample#3 is a few orders of magnitude smaller than that extracted from both sample#1 and sample#2, which are thicker crystals. This could be explained by invoking the following differences between these samples:

1) The energy gap decreases by increasing the number of layers, starting from a value of approximately 1.58 eV for the monolayer and slowly approaches the value of bulk material which is approximately 1.41 eV. However, this difference does not contribute significantly to the observed difference, because our beam is monochromatic and the photon energy is approximately 2.3 eV, or well above the gap for all samples.

2) The carrier recombination rate should be significantly larger for the thinner sample. Hence the back gate voltage, which separates charge carriers of opposite sign along the vertical direction, can more efficiently spatially separate carriers in thicker samples than in thinner ones. This will significantly decrease the PV current and PV efficiently in thinner samples, as experimentally seen.

3) In general, the Schottky barriers at both contacts are expected to be different among the various samples. This complicates the analysis of the PV response as a function of the number of atomic layers. We argued in this manuscript that the Schottky barriers in both electrodes act as two Schottky diodes which lead to the observed PV response in the absence of a fabricated PN junction.

## 4. Photocurrent and open circuit voltage as a function of applied optical power

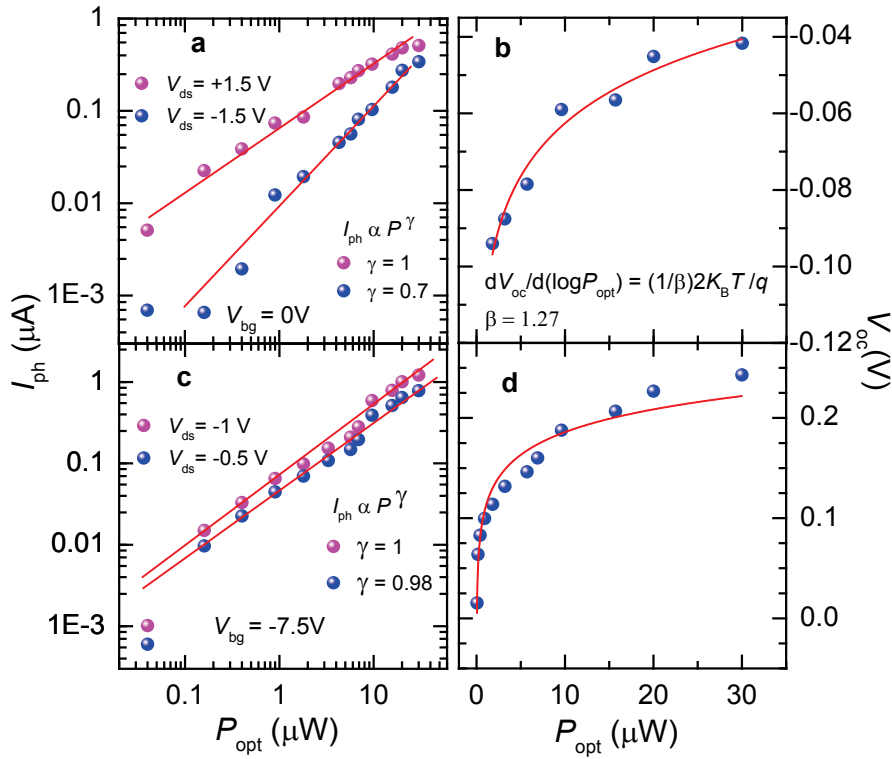

**Figure S5.** a) Photocurrent $I_{ph}$ as a function of the applied optical power $P_{opt}$ in a log-log scale, for two values of the bias voltage $V_{ds} = +1.5$ V (magenta markers) and $-1.5$ V (blue markers), respectively and zero gate voltage. b) Same as in a but under $V_{bg} = -7.5$ V and for $V_{ds} = -1$ (magenta markers) and $-0.5$ V (blue markers), respectively. In both figures red lines are linear fits. c) and d) $V_{oc}$ as a function of $P_{opt}$, for $V_{bg} = 0$ V and $V_{ds} = -1.5$ V, and for $V_{bg} = -7.5$ V and $V_{ds} = -0.5$ V, respectively. Red lines are logarithmic fits.

**Figure S5** above displays both photocurrent $I_{ph} = I_{ds}(P) - I_{ds}(P = 0\ W)$ and the extracted open circuit voltages $V_{oc}$ as functions of the applied laser power $P_{opt}$ for two values of the gate voltage $V_{bg}$ and for several values of the bias voltage. This data was collected from sample #1. As seen, $I_{ph}$ displays a power dependence on $P_{opt}$ or $I_{ph} \propto P^{\gamma}_{opt}$ with an exponent $\gamma$ ranging from 1 to 0.7. This suggests that the photo-thermoelectric effect might play a role in the observed photoresponse of our MoSe$_2$ field-effect transistors. $V_{oc}$ on the other hand displays the characteristic logarithmic dependence on $P_{opt}$, although the slope of the logarithmic fit, yield $\beta$ values closer to 1 which according to $dV_{OC}/d(\log P_{opt}) = 2k_B T / \beta e$ and as discussed in Ref. [S1] suggests the predominance of monomolecular recombination processes over the bimolecular or Langevin one (which would yield $\beta = 2$)

## 5. Photo responsivity and external quantum efficiencies

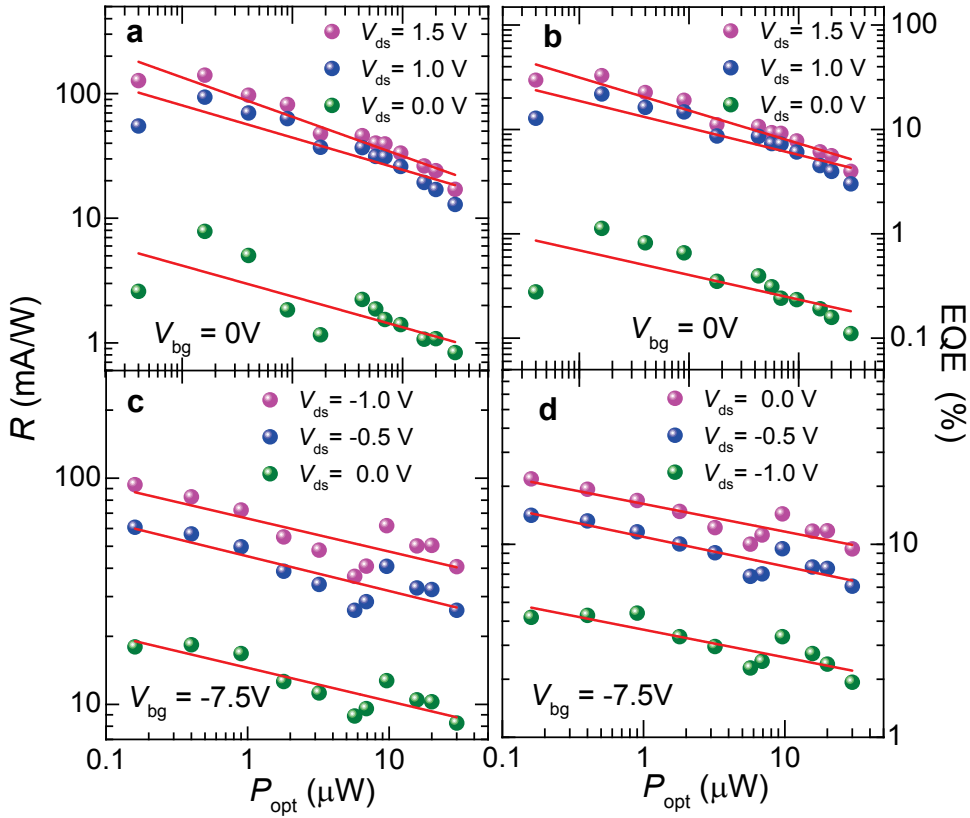

**Figure S6.** a) Photo responsivity $R = I_{ph}/P_{opt}$ as a function of $P_{opt}$ and for 3 values of the bias voltage, 0.0 (green markers), 1.0 (blue markers) and 1.5 V (magenta markers), respectively. Red lines are linear fits yielding nearly the same power law dependence $R \propto P^{\gamma}_{opt}$ with $\gamma \cong -0.3$. b) EQE as extracted from the data in a) yielding the same power dependence. c) Photo responsivity as a function of $P_{opt}$ under $V_{bg} = -7.5$ V and for 3 values of the bias voltage, 0.0 (green markers), 0.5 (blue markers) and -1.0 V (magenta markers), respectively. Red lines are linear fits yielding $\gamma \cong -0.15$. d) EQE from the data in c).

**Figure S6** above displays both the photoresponsivity $R = I_{ph}/P_{opt}$ and the external quantum efficiency as EQE $= hcR/e\lambda$, where $\lambda = 532$ nm, for device #1 and for several values of the bias and gate voltages. As seen, $R$ approaches ~100 mA/W at low $P_{opt}$ values under 0 as well as under an applied gate voltage. EQE on the other hand, approaches maximum values ranging between 20 and 30 % at low $P_{opt}$ values.

## 6. Values of the parameters of the Kirchoff-Shockley Circuit.

The two tables below present the values of the parameters $n_s$, $n_d$, $R_s$, $I_L^{(s)}$, $I_L^{(d)}$, $I_0^{(s)}$, and $I_0^{(d)}$ which enter in Eqs (1-4) of the Kirchoff-Shockley circuit and which we use to describe the fit to the experimental data.

Notice that by illuminating the devices, the only two parameters that we allow to change are the photocurrent parameters $I_L^{(s)}$, $I_L^{(d)}$, and the rest of the parameters remain the same as for the device in dark current. Also, notice that one of the main effects of the back gate voltage is to significantly alter the values of the Shockley parameters $I_0^{(s)}$, and $I_0^{(d)}$ where the role of the Schottky barrier is "hiding". These changes also cause the distribution of the photocurrent in terms of $I_L^{(s)}$, and $I_L^{(d)}$ to change.

| $V_{bg}$(V) | $P$ ($\mu$W) | $n_s$ | $n_d$ | $R_s$(M$\Omega$) | $I_L^{(s)}$ ($\mu$A) | $I_L^{(d)}$ ($\mu$A) | $I_D^{(s)}$ ($\mu$A) | $I_D^{(d)}$ ($\mu$A) |
|---|---|---|---|---|---|---|---|---|
| 0 | 0 | 1.3 | 1.1 | 0.02 | 0 | 0 | 0.012 | 0.0004 |
| 0 | 15.7 | 1.3 | 1.1 | 0.02 | 0.14 | 0.0 | 0.012 | 0.0004 |
| 0 | 30 | 1.3 | 1.1 | 0.02 | 0.28 | 0.0004 | 0.012 | 0.0004 |
| -7.5 | 0 | 1.744 | 1.1628 | 0.2 | 0 | 0 | 0.08 | 0.06 |
| -7.5 | 15.7 | 1.744 | 1.1628 | 0.2 | 0.1 | 0.88 | 0.08 | 0.06 |
| -7.5 | 30 | 1.744 | 1.1628 | 0.2 | 0.15 | 1.55 | 0.08 | 0.06 |

Table I: The value of the parameters for device #1 obtained by fitting the experimental results under various conditions.

| $V_{bg}$(V) | $P$ ($\mu$W) | $n_s$ | $n_d$ | $R_s$(M$\Omega$) | $I_L^{(s)}$ ($\mu$A) | $I_L^{(d)}$ ($\mu$A) | $I_D^{(s)}$ ($\mu$A) | $I_D^{(d)}$ ($\mu$A) |
|---|---|---|---|---|---|---|---|---|
| 0 | 0 | 1.1357 | 1.1628 | 1.2 | 0 | 0 | 0.00002 | 0.00001 |
| 0 | 15.7 | 1.1357 | 1.1628 | 1.2 | 0.45 | 0.16 | 0.00002 | 0.00001 |
| 0 | 30 | 1.20155 | 1.1628 | 0.5 | 0.48 | 0.32 | 0.00002 | 0.00001 |
| -7.5 | 0 | 1.2 | 1.0 | 0.02 | 0 | 0 | 0.01 | 0.001 |
| -7.5 | 15.7 | 1.2 | 1.0 | 0.02 | 0.55 | 0.01 | 0.01 | 0.001 |
| -7.5 | 30 | 1.2 | 1.0 | 0.02 | 1.2 | 0.04 | 0.01 | 0.001 |

Table II: The value of the parameters for device #2 obtained by fitting the experimental results under various conditions.

## 7. Evaluation of the photovoltaic response from dual Schottky diodes in ambipolar MoSe$_2$ field-effect transistors

Supplementary **Figure S6** presents an overall evaluation of the photovoltaic response of both MoSe$_2$ field-effect transistors whose data is discussed above. Figure 6a displays the short circuit current or Isc from sample #2, as well as the open circuit voltage $V_{oc}$ (magenta markers), both collected under $V_{bg} = 0$ V and as function of the illumination power density $p_{opt}$.

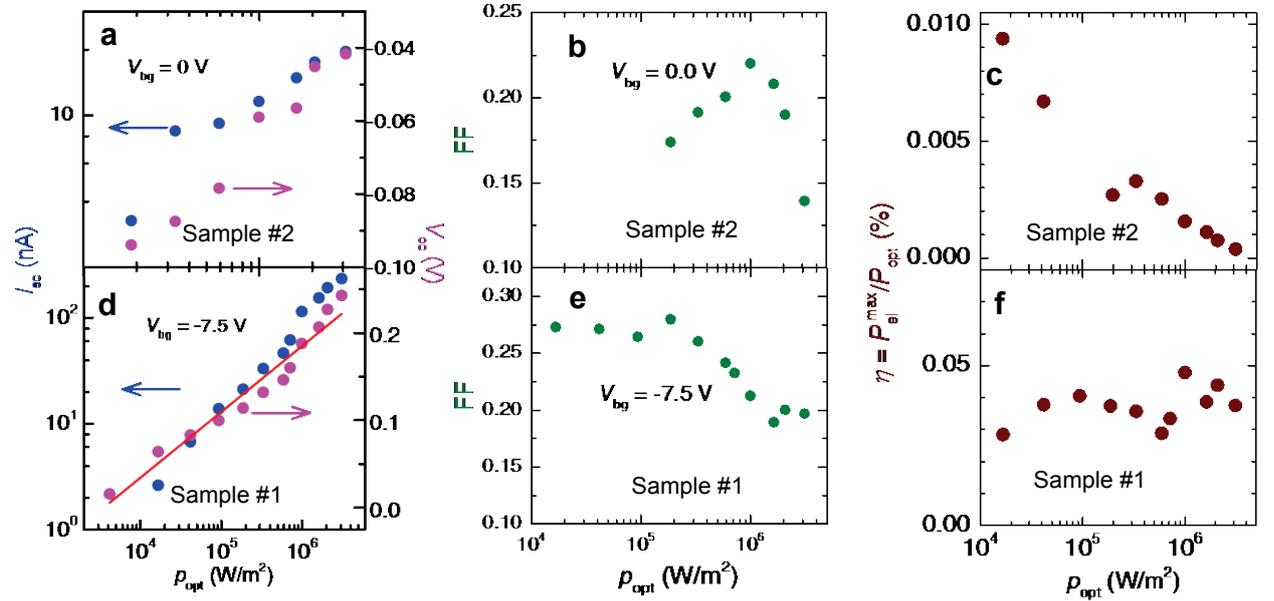

**Figure S7.** a) Short-circuit current $I_{sc} = I_{ds}$ ($V_{ds} = 0$ V) (blue markers) and open-circuit voltage $V_{oc} = V_{ds}(I_{ds} = 0$ A) (magenta markers) as functions of the applied illumination power density $p_{opt}$ under zero gate-voltage. b) Fill factor FF = $P_{el}^{max}$ / ($I_{sc}$ x $V_{oc}$) (green markers) as a function of $p_{opt}$. c) Photovoltaic efficiency $\eta$ defined here as $P_{el}^{max}/P_{opt}$, where $P_{opt}$ is total illumination power, as a function of $p_{opt}$ under $V_{bg} = 0$ V. d) Same as in a) but under $V_{bg} = -7.5$ V. Red line corresponds to a linear fit of $V_{oc}$ in a semi-log scale indicating that $V_{oc}$ displays a logarithmic dependence on $p_{opt}$. A linear fit of log($I_{sc}$) as a function log($p_{opt}$) yields an exponent $\alpha = 1.06$. e) Same as in b) but under $V_{bg} = -7.5$ V. f) Same as in c) but under $V_{bg} = -7.5$ V.

Figure S7b presents the photovoltaic fill factor FF = $P_{el}^{max}/(J_{sc} \times V_{oc})$ as a function of $p_{opt}$. Figure S7c presents the photovoltaic efficiency following a commonly used convention $\eta = P_{el}^{max}/P_{opt}$, where $P_{opt}$ is the total illumination power. One extracts $\eta$ values in the order of just 1 x 10$^{-2}$ % at low $p_{opt}$ suggesting a mild gradient of chemical potential which separates electron-hole pairs and allows their collection at the contacts before recombination. Figure S7d presents similar data to the one displayed in Figure 6a but for sample #1 under a gate voltage $V_{bg}$ = - 7.5 V. The red line corresponds to a semi-logarithmic fit of $V_{oc}$ as a function of $p_{opt}$. Figure S6e presents similar data to one displayed in Figure 7b but for sample #1 under a gate voltage $V_{bg}$ = - 7.5 V. By comparing with Figure S7b notice how one extracts larger FF values, particularly at low $p_{opt}$. Finally, Figure S6f indicates that one would extract photovoltaic efficiencies approaching ~ 0.05 % at low power densities in contrast to Figure 5c which yields $\eta \cong 0.01$ %. These $\eta$ values certainly are far smaller than those extracted by Refs. S2-S3 which report photovoltaic efficiencies ranging from 1.8 to 5.23 %. Nevertheless, we used the Ti:Au for both contacts; notice that the choice of two distinct metals for each contact such as Au and Pd was reported to yield conversion efficiencies approaching 2.5 %.[S4] Our results suggest that a quite simple architecture as the one discussed here, but with two metals having quite distinct work functions such as Pd and Sc (or Sm), should yield even higher photovoltaic conversion efficiencies.

# Optoelectronic switch based on intrinsic dual Schottky diodes in ambipolar MoSe$_2$ field-effect transistors

*Nihar R. Pradhan,*[*] *Zhengguang Lu*, *Daniel Rhodes*, *Dmitry Smirnov*, *Efstratios Manousakis*,[*]

*and Luis Balicas*[*]

Dr. N. R. Pradhan, Z. Lu, D. Rhodes, Dr. D. Smirnov, Prof. E. Manousakis, Dr. L. Balicas
National High Magnetic Field Lab, Florida State University, 1800 E. Paul Dirac Drive, Tallahassee, Florida 32310, United States.
E-mail: balicas@magnet.fsu.edu,
E-Mail: pradhan@magnet.fsu.edu

Prof. E. Manousakis
Florida State University, Department of Physics, Tallahassee, FL 32306, United States,

Prof. E. Manousakis
Department of Physics, University of Athens, Panepistimioupolis, Zografos, GR-157 84 Athens, Greece.

**8. Current as a function of bias voltage characteristics for both MoSe$_2$ field effect transistors.**

**Figure S1** displays the *I-V* characteristics of samples #1 and #2 under several back voltages and under dark conditions, revealing a pronounced non-linear *I-V* response from sample #1, particularly at low excitation voltages. This non-ohmic behavior can be attributed to pronounced Schottky barriers between the MoSe$_2$ channel and the Au:Ti contacts. In contrast, sample #2 presents a nearly linear, or ohmic like response at low bias voltages. This indicates that thermionic emission can promote charge carriers across the barriers at room temperature.

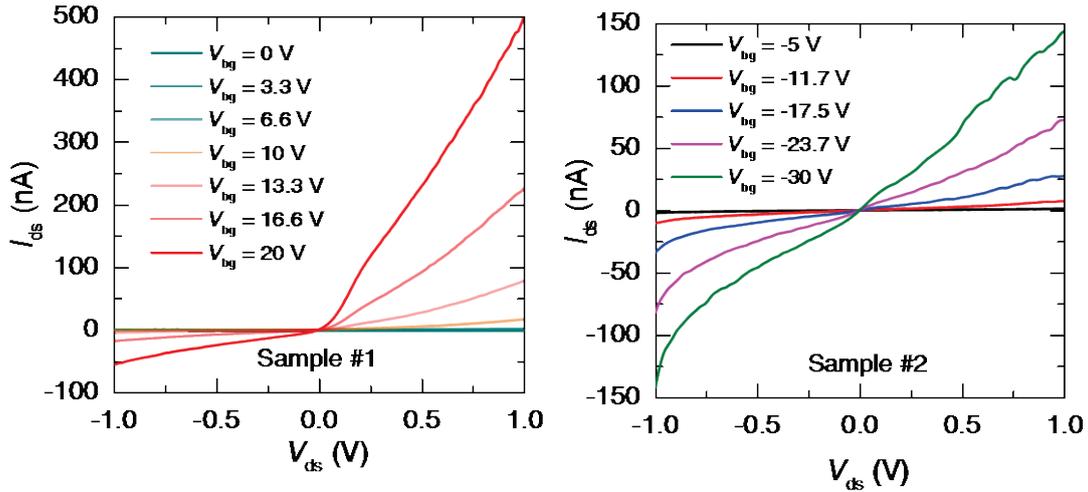

**Figure S1.** Left panel: Drain to source current $I_{ds}$ as a function of the bias voltage $V_{ds}$ for sample #1 and for several values of back gate voltage $V_{bg}$. Right panel: same as in the left panel but for sample #2.

## 9. Experimental set-up for measuring the electrical response under illumination power.

**Figure S2** below presents a sketch of the experimental set up used for the electrical characterization of our samples under dark or illumination conditions. The diode laser beam is sent to the sample through an optical fiber, subsequently collimated, and focused through an achromatic objective. If necessary its intensity can be modulated by using neutral density filters. The sample is measured through a dual channel sourcemeter.

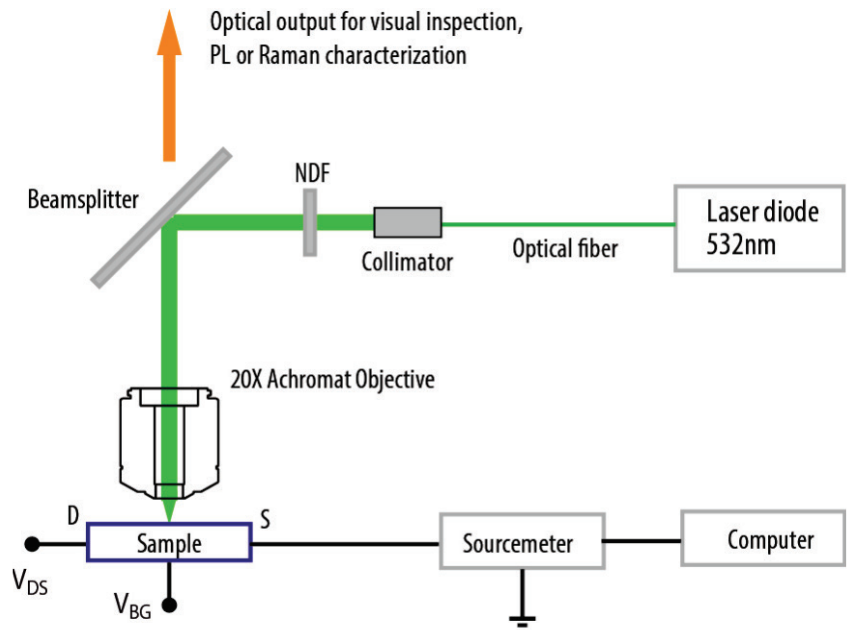

**Figure S2.** Schematics of the experimental set-up used to measure the electrical response of our MoSe$_2$ transistors under optical stimulation.

## 10. Light induced rectification response in a second multi-layered and a tri-layered MoSe$_2$ field-effect transistor

**Figure S3** below displays the overall photo response from a second MoSe$_2$ field-effect transistor (sample #2 in the main text) under illumination by a λ = 532 nm laser with a spot size of 3.5 μm.

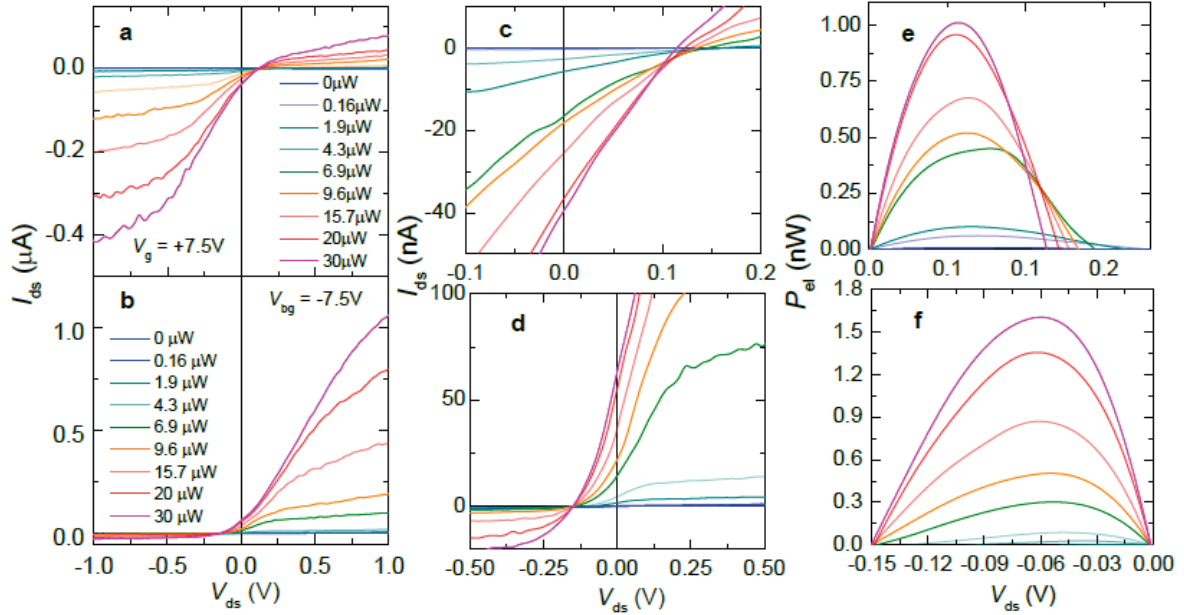

**Figure S3.** a) Drain to source current $I_{ds}$ as a function of the bias voltage $V_{ds}$ for several values of the illumination power and under a gate voltage $V_{bg}$ = - 7.5 V. b) Same as in a) but under $V_{bg}$ = - 7.5 V. The line color scheme which associated to the laser power is the same for all figures. c) Same as in a) but in an amplified scale. d) Same as in b) but in an amplified scale. Notice, by comparing c) and d), how under both gate voltages one obtains relatively similar values for the short circuit currents $I_{sc}$ = $I_{ds}(V_{ds} = 0$ V). e) Photo-generated electrical power $P_{el} = I_{ds}$ x $V_{ds}$ as extracted from the fourth quadrant, spanning from $V_{ds}$ = 0 V to $I_{ds}$ = 0 A, in c). f) same as in e) but from the data in d). Notice how in sharp contrast to sample #1, sample # 2 yields comparable $P_{el}$ values for both gate voltages although this higher degree of symmetry leads to much lower photogenerated power values when compared to the ones extracted from sample #1 under $V_{bg}$ = - 7.5 V.

Notice how for this sample by inverting the sign of the gate voltage one can also switch the sense of current rectification. Although, with respect to sample #1, in this sample the sense of current rectification displays an opposite dependence on the sign of the gate voltage. For sample #1 one extracts a sizeable $I_{ds}$ only when *both* $V_{ds}$ and $V_{bg}$ are either > 0 or < 0 in contrast to what is seen here. Again, this indicates that the gate voltage is modulating the amplitude of each Schottky barrier in an independent and non-trivial manner, perhaps by pinning the Fermi level at distinct impurity levels at each contact. Notice that for sample #2 under zero bias and for both gate voltages one extracts similar values for the short circuit current suggesting that for this sample the gate voltage can invert the assymmetry between both Schottky barriers. It also points towards a lower degree of asymmetry in the height of both barriers. As seen, a higher degree of symmetry leads to lower electrical power conversion levels, when compared for example, to sample #1 under $V_{bg}$ = - 7.5 V (see Figure 4 f in the main text). For the sake of comparison with the data displayed in the main text (from a multi-layered sample) in **Figure S4** below we display data from a tri-layered sample, or sample #3, indicating that the rectification effect is also observed in thinner samples characterized by a larger, and nearly direct semiconducting gap.

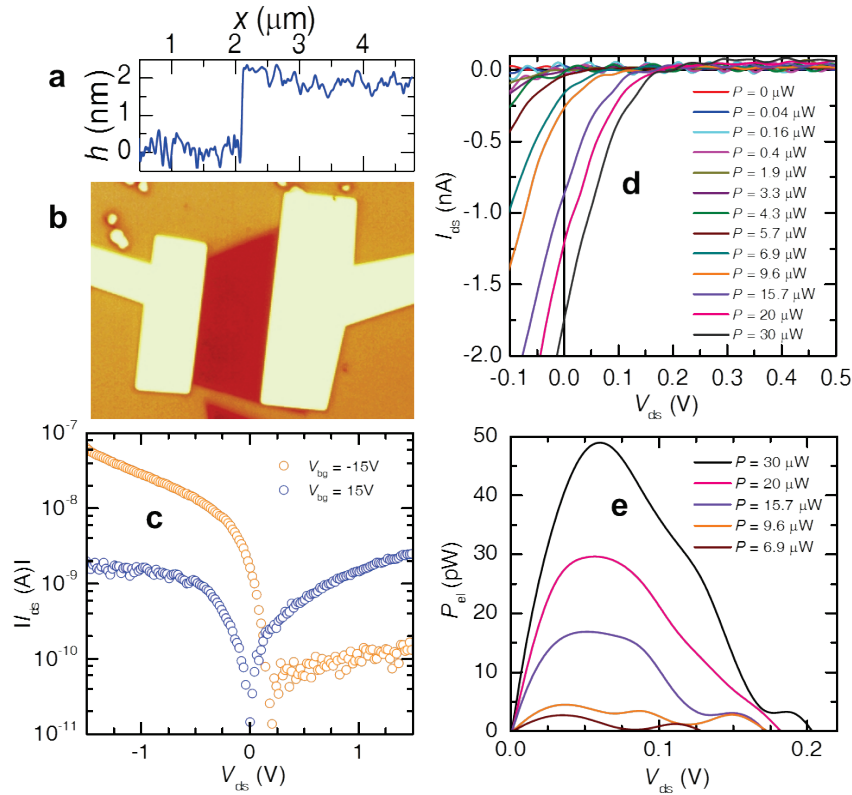

**Figure S4.** a) Atomic force microscopy height profile of a third sample, i.e. sample #3, indicating a thickness of approximately 2 nm or 3 atomic layers. b) Micrograph of sample#3. c) Drain to source current $I_{ds}$ as a function of the bias voltage $V_{ds}$ for two values of the gate voltage, i.e. respectively 15 and -15 V under an illumination power of $P$ = 30 µW. d) $I_{ds}$ as a function of $V_{ds}$ for several values of $P$. Notice the finite photogenerated current in absence of bias voltage. e) Photogenerated electrical power $P_{el} = I_{ds} \times V_{ds}$ as a function $V_{ds}$ for several values of the illumination power.

As seen in Figure S4e the maximum $P_{el}$ output power and more generally the response of sample#3 is a few orders of magnitude smaller than that extracted from both sample#1 and sample#2, which are thicker crystals. This could be explained by invoking the following differences between these samples:

1) The energy gap decreases by increasing the number of layers, starting from a value of approximately 1.58 eV for the monolayer and slowly approaches the value of bulk material which is approximately 1.41 eV. However, this difference does not contribute significantly to the observed difference, because our beam is monochromatic and the photon energy is approximately 2.3 eV, or well above the gap for all samples.

2) The carrier recombination rate should be significantly larger for the thinner sample. Hence the back gate voltage, which separates charge carriers of opposite sign along the vertical direction, can more efficiently spatially separate carriers in thicker samples than in thinner ones. This will significantly decrease the PV current and PV efficiently in thinner samples, as experimentally seen.

3) In general, the Schottky barriers at both contacts are expected to be different among the various samples. This complicates the analysis of the PV response as a function of the number of atomic layers. We argued in this manuscript that the Schottky barriers in both electrodes act as two Schottky diodes which lead to the observed PV response in the absence of a fabricated PN junction.

## 11. Photocurrent and open circuit voltage as a function of applied optical power

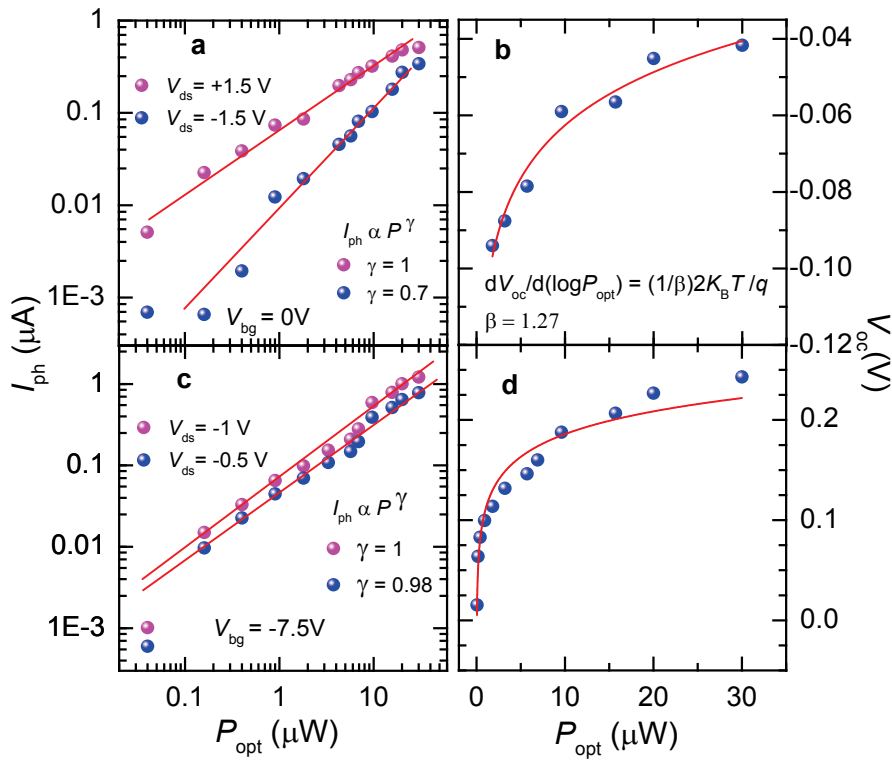

**Figure S5.** a) Photocurrent $I_{ph}$ as a function of the applied optical power $P_{opt}$ in a log-log scale, for two values of the bias voltage $V_{ds}$ = +1.5 V (magenta markers) and -1.5 V (blue markers), respectively and zero gate voltage. b) Same as in a but under $V_{bg}$ = - 7.5 V and for $V_{ds}$ = -1 (magenta markers) and -0.5 V (blue markers), respectively. In both figures red lines are linear fits. c) and d) $V_{oc}$ as a function of $P_{opt}$, for $V_{bg}$ = 0 V and $V_{ds}$ = -1.5 V, and for $V_{bg}$ = - 7.5 V and $V_{ds}$ = - 0.5 V, respectively. Red lines are logarithmic fits.

**Figure S5** above displays both photocurrent $I_{ph} = I_{ds}(P) - I_{ds}(P = 0\text{ W})$ and the extracted open circuit voltages $V_{oc}$ as functions of the applied laser power $P_{opt}$ for two values of the gate voltage $V_{bg}$ and for several values of the bias voltage. This data was collected from sample #1. As seen, $I_{ph}$ displays a power dependence on $P_{opt}$ or $I_{ph} \propto P_{opt}^{\gamma}$ with an exponent $\gamma$ ranging from 1 to 0.7. This suggests that the photo-thermoelectric effect might play a role in the observed photoresponse of our MoSe$_2$ field-effect transistors. $V_{oc}$ on the other hand displays the characteristic logarithmic dependence on $P_{opt}$, although the slope of the logarithmic fit, yield $\beta$ values closer to 1 which according to $dV_{OC}/d(\log P_{opt}) = 2k_B T / \beta e$ and as discussed in Ref. [S1] suggests the predominance of monomolecular recombination processes over the bimolecular or Langevin one (which would yield $\beta = 2$)

## 12. Photo responsivity and external quantum efficiencies

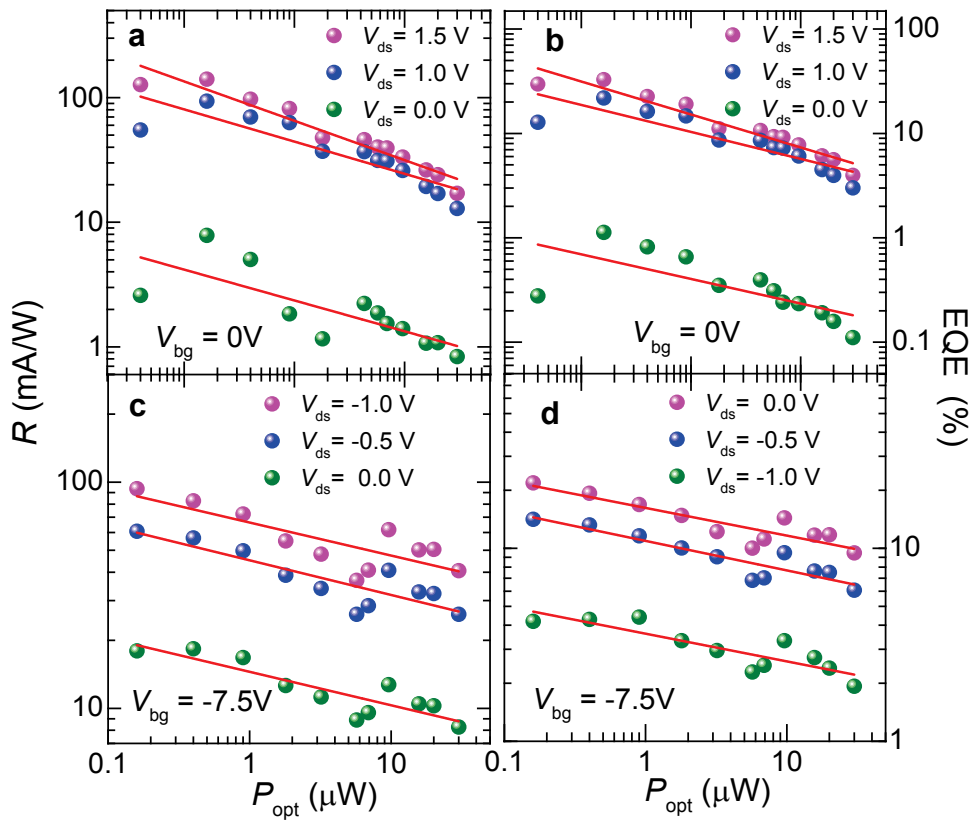

**Figure S6.** a) Photo responsivity $R = I_{ph}/P_{opt}$ as a function of $P_{opt}$ and for 3 values of the bias voltage, 0.0 (green markers), 1.0 (blue markers) and 1.5 V(magenta markers), respectively. Red lines are linear fits yielding nearly the same power law dependence $R \propto P_{opt}^{\gamma}$ with $\gamma \cong -0.3$. b) EQE as extracted from the data in a) yielding the same power dependence. c) Photo responsivity as a function of $P_{opt}$ under $V_{bg} = -7.5$ V and for 3 values of the bias voltage, 0.0 (green markers), 0.5 (blue markers) and -1.0 V (magenta markers), respectively. Red lines are linear fits yielding $\gamma \cong -0.15$. d) EQE from the data in c).

**Figure S6** above displays both the photoresponsivity $R = I_{ph}/P_{opt}$ and the external quantum efficiency as EQE $= hcR/e\lambda$, where $\lambda = 532$ nm, for device #1 and for several values of the bias and gate voltages. As seen, $R$ approaches ~100 mA/W at low $P_{opt}$ values under 0 as well as under an applied gate voltage. EQE on the other hand, approaches maximum values ranging between 20 and 30 % at low $P_{opt}$ values.

### 13. Values of the parameters of the Kirchoff-Shockley Circuit.

The two tables below present the values of the parameters $n_s$, $n_d$, $R_s$, $I_L^{(s)}$, $I_L^{(d)}$, $I_0^{(s)}$, and $I_0^{(d)}$ which enter in Eqs (1-4) of the Kirchoff-Shockley circuit and which we use to describe the fit to the experimental data.

Notice that by illuminating the devices, the only two parameters that we allow to change are the photocurrent parameters $I_L^{(s)}$, $I_L^{(d)}$, and the rest of the parameters remain the same as for the device in dark current. Also, notice that one of the main effects of the back gate voltage is to significantly alter the values of the Shockley parameters $I_0^{(s)}$, and $I_0^{(d)}$ where the role of the Schottky barrier is "hiding". These changes also cause the distribution of the photocurrent in terms of $I_L^{(s)}$, and $I_L^{(d)}$ to change.

| $V_{bg}$(V) | $P$ ($\mu$W) | $n_s$ | $n_d$ | $R_s$(M$\Omega$) | $I_L^{(s)}$ ($\mu$A) | $I_L^{(d)}$ ($\mu$A) | $I_D^{(s)}$ ($\mu$A) | $I_D^{(d)}$ ($\mu$A) |
|---|---|---|---|---|---|---|---|---|
| 0 | 0 | 1.3 | 1.1 | 0.02 | 0 | 0 | 0.012 | 0.0004 |
| 0 | 15.7 | 1.3 | 1.1 | 0.02 | 0.14 | 0.0 | 0.012 | 0.0004 |
| 0 | 30 | 1.3 | 1.1 | 0.02 | 0.28 | 0.0004 | 0.012 | 0.0004 |
| -7.5 | 0 | 1.744 | 1.1628 | 0.2 | 0 | 0 | 0.08 | 0.06 |
| -7.5 | 15.7 | 1.744 | 1.1628 | 0.2 | 0.1 | 0.88 | 0.08 | 0.06 |
| -7.5 | 30 | 1.744 | 1.1628 | 0.2 | 0.15 | 1.55 | 0.08 | 0.06 |

Table I: The value of the parameters for device #1 obtained by fitting the experimental results under various conditions.

| $V_{bg}$(V) | $P$ ($\mu$W) | $n_s$ | $n_d$ | $R_s$(M$\Omega$) | $I_L^{(s)}$ ($\mu$A) | $I_L^{(d)}$ ($\mu$A) | $I_D^{(s)}$ ($\mu$A) | $I_D^{(d)}$ ($\mu$A) |
|---|---|---|---|---|---|---|---|---|
| 0 | 0 | 1.1357 | 1.1628 | 1.2 | 0 | 0 | 0.00002 | 0.00001 |
| 0 | 15.7 | 1.1357 | 1.1628 | 1.2 | 0.45 | 0.16 | 0.00002 | 0.00001 |
| 0 | 30 | 1.20155 | 1.1628 | 0.5 | 0.48 | 0.32 | 0.00002 | 0.00001 |
| -7.5 | 0 | 1.2 | 1.0 | 0.02 | 0 | 0 | 0.01 | 0.001 |
| -7.5 | 15.7 | 1.2 | 1.0 | 0.02 | 0.55 | 0.01 | 0.01 | 0.001 |
| -7.5 | 30 | 1.2 | 1.0 | 0.02 | 1.2 | 0.04 | 0.01 | 0.001 |

Table II: The value of the parameters for device #2 obtained by fitting the experimental results under various conditions.

## 14. Evaluation of the photovoltaic response from dual Schottky diodes in ambipolar MoSe$_2$ field-effect transistors

Supplementary **Figure S6** presents an overall evaluation of the photovoltaic response of both MoSe$_2$ field-effect transistors whose data is discussed above. Figure 6a displays the short circuit current or Isc from sample #2, as well as the open circuit voltage $V_{oc}$ (magenta markers), both collected under $V_{bg}$ = 0 V and as function of the illumination power density $p_{opt}$.

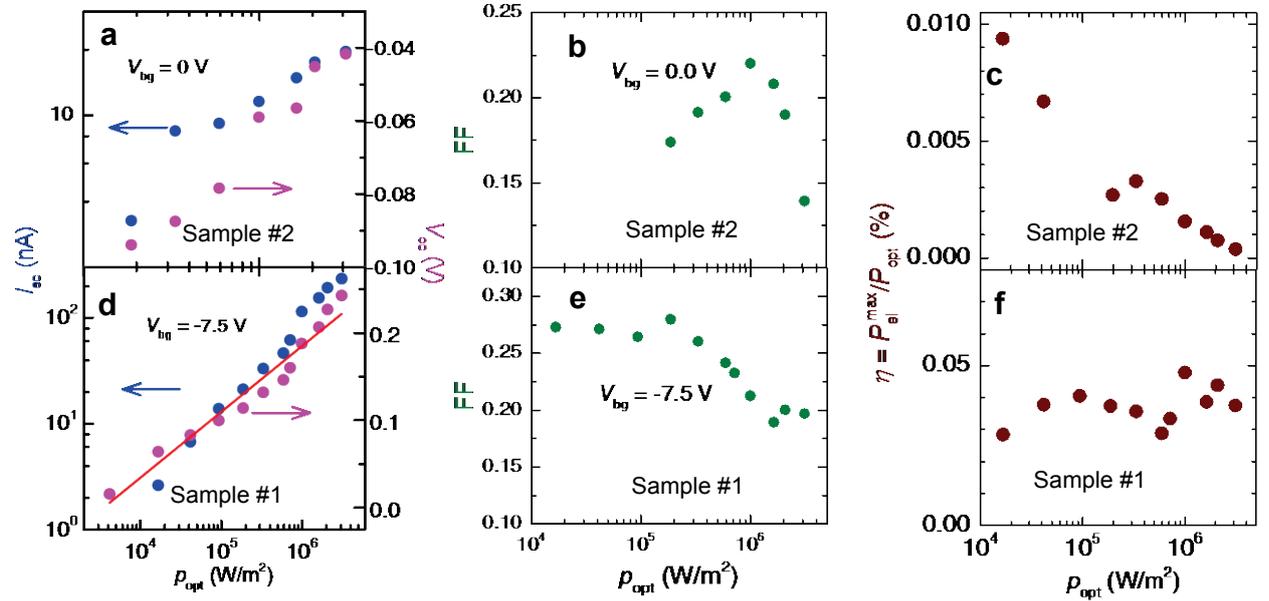

**Figure S7.** a) Short-circuit current $I_{sc} = I_{ds}$ ($V_{ds}$ = 0 V) (blue markers) and open-circuit voltage $V_{oc} = V_{ds}(I_{ds} = 0$ A) (magenta markers) as functions of the applied illumination power density $p_{opt}$ under zero gate-voltage. b) Fill factor FF = $P_{el}^{max}$ / ($I_{sc}$ x $V_{oc}$) (green markers) as a function of $p_{opt}$. c) Photovoltaic efficiency $\eta$ defined here as $P_{el}^{max}/P_{opt}$, where $P_{opt}$ is total illumination power, as a function of $p_{opt}$ under $V_{bg}$ = 0 V. d) Same as in a) but under $V_{bg}$ = - 7.5 V. Red line corresponds to a linear fit of $V_{oc}$ in a semi-log scale indicating that $V_{oc}$ displays a logarithmic dependence on $p_{opt}$. A linear fit of log($I_{sc}$) as a function log($p_{opt}$) yields an exponent α = 1.06. e) Same as in b) but under $V_{bg}$ = - 7.5 V. f) Same as in c) but under $V_{bg}$ = - 7.5 V.

Figure S7b presents the photovoltaic fill factor FF = $P_{el}^{max}/(J_{sc} \times V_{oc})$ as a function of $p_{opt}$. Figure S7c presents the photovoltaic efficiency following a commonly used convention $\eta = P_{el}^{max}/P_{opt}$, where $P_{opt}$ is the total illumination power. One extracts $\eta$ values in the order of just $1 \times 10^{-2}$ % at low $p_{opt}$ suggesting a mild gradient of chemical potential which separates electron-hole pairs and allows their collection at the contacts before recombination. Figure S7d presents similar data to the one displayed in Figure 6a but for sample #1 under a gate voltage $V_{bg}$ = - 7.5 V. The red line corresponds to a semi-logarithmic fit of $V_{oc}$ as a function of $p_{opt}$. Figure S6e presents similar data to one displayed in Figure 7b but for sample #1 under a gate voltage $V_{bg}$ = - 7.5 V. By comparing with Figure S7b notice how one extracts larger FF values, particularly at low $p_{opt}$. Finally, Figure S6f indicates that one would extract photovoltaic efficiencies approaching ~ 0.05 % at low power densities in contrast to Figure 5c which yields $\eta \cong 0.01$ %. These $\eta$ values certainly are far smaller than those extracted by Refs. S2-S3 which report photovoltaic efficiencies ranging from 1.8 to 5.23 %. Nevertheless, we used the Ti:Au for both contacts; notice that the choice of two distinct metals for each contact such as Au and Pd was reported to yield conversion efficiencies approaching 2.5 %.[S4] Our results suggest that a quite simple architecture as the one discussed here, but with two metals having quite distinct work functions such as Pd and Sc (or Sm), should yield even higher photovoltaic conversion efficiencies.